\documentclass{article}
\usepackage{iclr2027_conference,times}

\usepackage{amsmath,amsfonts,bm}

\def\eqref#1{equation~\ref{#1}}

\def\1{\bm{1}}

\DeclareMathAlphabet{\mathsfit}{\encodingdefault}{\sfdefault}{m}{sl}
\SetMathAlphabet{\mathsfit}{bold}{\encodingdefault}{\sfdefault}{bx}{n}

\usepackage[utf8]{inputenc} 
\usepackage[T1]{fontenc}    
\usepackage{hyperref}       
\usepackage{url}            
\usepackage{booktabs}       
\usepackage{amsfonts}       
\usepackage{nicefrac}       
\usepackage{microtype}      
\usepackage{xcolor}         

\usepackage{algorithm}
\usepackage{algorithmic}
\usepackage{amsmath}
\usepackage{multirow}
\usepackage{makecell}
\usepackage{enumitem}
\usepackage{subcaption}
\usepackage{graphicx}

\title{PRiSE-EEG: A Prior-Guided Foundation Model with Depth-Stratified Experts for Cross-Paradigm EEG Representation Learning}

\iclrfinalcopy
\author{%
  Wei Xiong$^{1}$, Jiangtong Li$^{1}$, Kun Zhu$^{1}$,
  Jie Li$^{1,2}$, Changjun Jiang$^{1}$ \\
  $^{1}$School of Computer Science and Technology Tongji University, Shanghai, China \\
  $^{2}$Translational Research Center, Shanghai Yangzhi Rehabilitation Hospital \\
  (Shanghai Sunshine Rehabilitation Center), Tongji University, Shanghai, China \\
  \texttt{\{xw1216, jiangtongli\}@tongji.edu.cn}
}

\begin{document}

\maketitle
\fancyhead{}

\begin{abstract}
EEG foundation models aim to learn reusable representations across heterogeneous paradigms, yet existing approaches often use uniform adaptation mechanisms and are typically reported under separate downstream fine-tuning protocols. In this work, we first analyze dense EEG Transformers from two complementary perspectives. Gradient similarity across six downstream datasets reveals substantial optimization conflicts among EEG paradigms, while CKA analysis on mixed-paradigm batches shows a consistent depth-wise transition: shallow layers preserve stronger cross-paradigm similarity, whereas deeper layers become increasingly specialized. Motivated by these findings, we propose \textbf{PRiSE-EEG}, a prior-guided EEG foundation model with CKA-calibrated Depth-Stratified Experts. PRiSE-EEG forms continuous multi-channel EEG patches using weak static cortical and network priors and dynamic short-time channel interactions, then allocates shared and specialized experts across MoE Transformer blocks according to a sigmoid mapping from layer-wise CKA sharedness. This design preserves common EEG regularities in early blocks while assigning more specialized capacity to later task-specific transformations. Experiments on 12 public EEG benchmarks show strong performance. Compact ablations further show that CKA-derived expert allocation improves over dense Transformers, uniform MoE, and manually fixed shared-specific expert ratios. Code is available in the \href{https://github.com/xw1216/PRiSE-EEG}{GitHub repository}.
\end{abstract}

\section{Introduction}\label{sec:intro}

\begin{figure}[t]
\centering
\includegraphics[width=\linewidth]{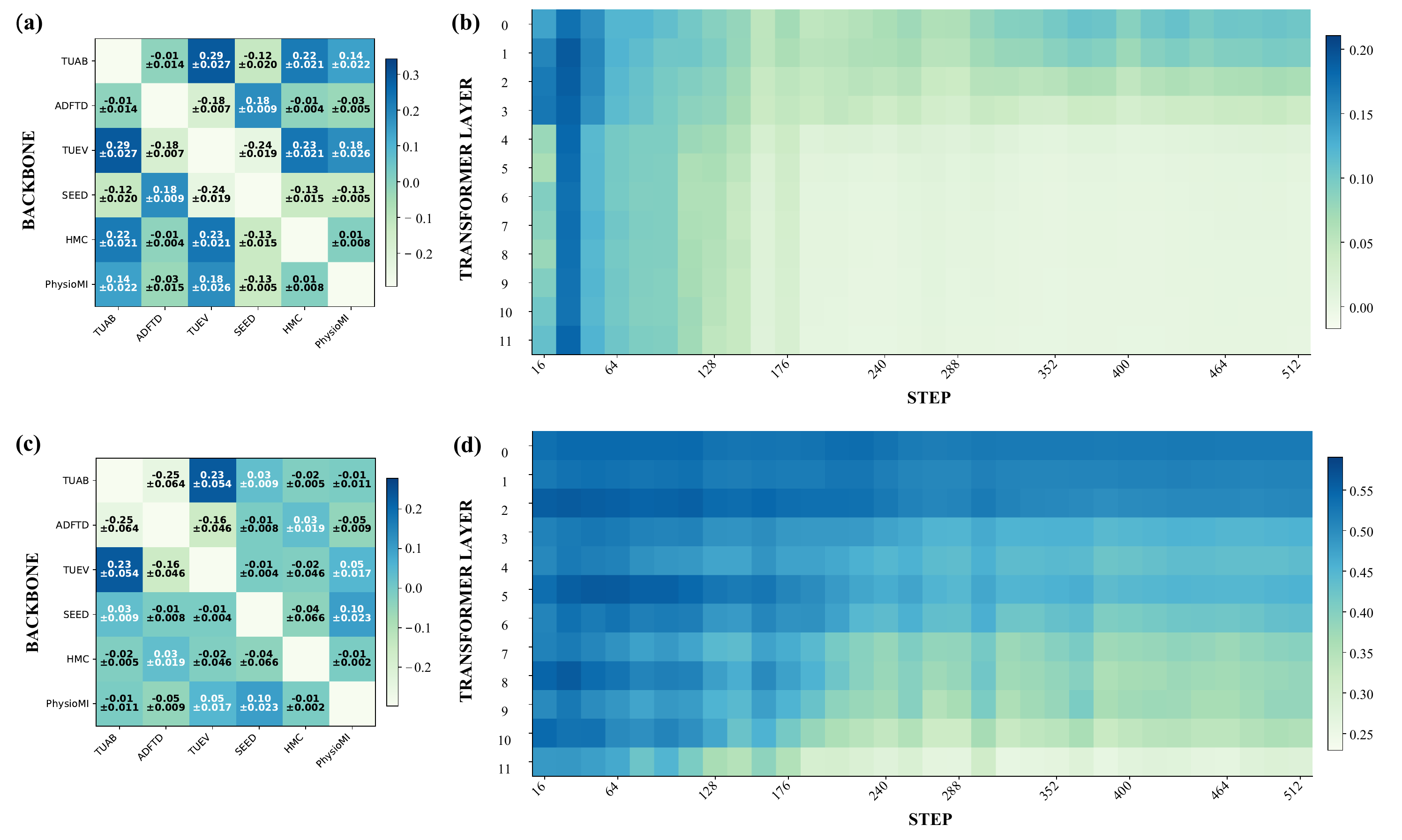}
\caption{
Pairwise cosine similarity of downstream fine-tuning gradients across datasets for LaBraM (a) and CSBrain (c), respectively. Lower similarities indicate task-dependent optimization conflicts. Layer-wise sample CKA across pretraining steps for LaBraM (b) and CSBrain (d), respectively. Shallow layers show stronger cross-paradigm alignment, whereas deeper layers become more specialized, motivating CKA-calibrated depth-stratified expert allocation.
}
\label{fig:motivation}
\end{figure}

Electroencephalography (EEG) signals differ substantially in montage, temporal statistics, dominant rhythms, label semantics, and task-relevant spatial patterns. Classical models for EEG suffer from subject-, session-, and dataset-specific variability~\citep{lawhern2018eegnet, song2022eegconformer,demir2021eeg}. EEG foundation models (EEG-FM) aim to learn reusable representations across heterogeneous paradigms, yet cross-paradigm transfer remains difficult. Therefore, an EEG-FM must preserve general EEG regularities while still adapting to task-specific evidence. Existing models usually address this problem either by improving pretraining and tokenization, or by adding adaptation mechanisms during fine-tuning. Recent works such as BENDR~\citep{kostas2021bendr}, BIOT~\citep{yang2023biot}, EEGPT~\citep{wang2024eegpt}, CBraMod~\citep{wang2024cbramod}, CSBrain~\citep{zhou2025csbrain} improve scalable self-supervised pretraining and heterogeneous modeling. Discrete-token approaches, including LaBraM~\citep{jiang2024large}, CodeBrain~\citep{ma2025codebrain} and TFM-Tokenizer~\citep{pradeepkumar2025tfm-tokenize}, further show that how EEG is converted into discrete tokens is critical for representation quality. In parallel, EEGMoE~\citep{gao2026eegmoe} demonstrates that domain-specific experts can help decouple heterogeneous EEG domains in pre-training stage. 
However, two issues remain insufficiently explored. 
First, existing EEG-FMs rarely model cross-task relationships, although heterogeneous paradigms may induce different update directions. 
Second, spatial structure beyond electrode coordinates or montage embeddings remains underused for patch formation.

Although sensor-space EEG does not provide direct source-level measurements, it still preserves structured spatial regularities that are useful for representation learning. 
Electrophysiological studies show that scalp recordings contain coarse network-level organization~\citep{brookes2011investigating, hipp2012large, xavier2025consistency}, while EEG microstates and lag-sensitive coupling analyses suggest that multi-channel interactions can change within sub-100 ms windows~\citep{vicente2008dynamical, michel2018microstates, van2010microstate, mehra2025zero}. 
These findings suggest that more sensor-space prior and short-window dynamics can provide weak guidance for EEG learning.

We explore the task-relation issue afterwards by examining whether heterogeneous EEG tasks impose compatible updates at fine-tuning stage, as shown in Fig.~\ref{fig:motivation}.
We train dense Transformer baselines on six downstream datasets and compute pairwise gradient similarities. The resulting heatmaps show clear optimization-direction discrepancies across tasks, suggesting that fully shared dense parameters may face conflicting update preferences across paradigms~\citep{sener2018multi}. This motivates conditional capacity through MoE, where different inputs can activate different expert parameters. However, a uniform MoE allocation still ignores where shared and specialized capacity should appear across model depth. We therefore examine the representation geometry of a dense pilot EEG Transformer using Centered Kernel Alignment (CKA)~\citep{kornblith2019similarity}. Given mixed-paradigm batches, we compute block-wise CKA across predefined pretraining checkpoints, with within-paradigm and between-paradigm similarities averaged separately. The analysis shows a consistent depth-wise pattern: early blocks retain higher cross-paradigm similarity, while later blocks become increasingly task- or paradigm-specific. This suggests that EEG representations follow a general-to-specific transition across the pretraining Transformer stack, and that shared and specialized experts should not be allocated uniformly across all MoE blocks.

Guided by these empirical observations, we propose PRiSE-EEG, a Prior-guided EEG foundation model with depth-Stratified Experts. PRiSE-EEG forms continuous multi-channel EEG patches by combining weak static region and network priors with dynamic short-window channel interactions. On top of these patches, PRiSE-EEG introduces a CKA-calibrated expert allocation rule. For each MoE Transformer block, we estimate the ratio between between-paradigm and within-paradigm CKA, and map it to the shared expert proportion. Blocks with higher sharedness receive more shared experts, while blocks with lower sharedness allocate more capacity to specialized experts.

We evaluate PRiSE-EEG on 12 public EEG benchmarks covering sleep staging, motor imagery, emotion recognition, and other cognitive or clinical tasks. In evaluations across multiple EEG paradigms, PRiSE-EEG outperforms dense Transformer baselines, uniform Mixture-of-Expert(MoE) variants, and fixed-ratio shared-specific expert allocations. These results show that expert specialization in EEG foundation models should not only be input-adaptive, but also depth-stratified according to the representation structure learned by dense pilot models. Our contributions are summarized as follows:
\begin{itemize}[topsep=4pt, itemsep=-1pt]
    \item We characterize two issues in cross-paradigm EEG learning, including conflicting task updates across layers and underused weak sensor-space structure.
    \item We propose CKA-calibrated Depth-Stratified Experts, which convert block-wise CKA sharedness into shared expert proportions through a simple sigmoid allocation rule.
    \item We introduce PRiSE-EEG, combining prior-guided continuous tokenization with depth-stratified expert allocation, and validate it on 12 EEG benchmarks.
\end{itemize}

\section{Related Works}\label{sec:related-works}
\subsection{EEG Foundation Models and Tokenization}
Early convolutional, graph-based and attention-based models improve EEG spatial-temporal feature extraction but are commonly optimized for specific subjects or datasets~\citep{lawhern2018eegnet,song2022eegconformer,demir2021eeg,subhrajit2019chrono,song2021transformer}. Recent EEG foundation models learn transferable representations through self-supervised pretraining on heterogeneous EEG corpora, including BENDR~\citep{kostas2021bendr}, BIOT~\citep{yang2023biot}, EEGPT~\citep{wang2024eegpt}, CBraMod~\citep{wang2024cbramod}, CSBrain~\citep{zhou2025csbrain}, and REVE~\citep{Ouahidi2025reve}. Tokenization has also become central: LaBraM~\citep{jiang2024large} learns vector-quantized neural codes, CodeBrain~\citep{ma2025codebrain} decouples temporal and frequency tokens, and TFM-Tokenizer~\citep{pradeepkumar2025tfm-tokenize} learns single-channel time-frequency motifs. MMM organizes EEG spatial information through a unified topology and a multi-level channel hierarchy~\citep{yi2023mmm}. PRiSE-EEG combines learnable static spatial biases with sample-dependent short-time dynamic groups to form continuous multi-channel representations, and further uses layer-wise representation sharedness to allocate shared and specialized experts.

\subsection{Conditional Experts and Depth-wise Representation Analysis}
MoE models provide scalable conditional computation through sparsely gated experts~\citep{shazeer2017outrageously,fedus2022switch}, and have also been used in neuroimaging for heterogeneous aging modeling~\citep{eavani2016capturing} and cross-subject fMRI decoding~\citep{wei2025more}. In EEG representation learning, EEGMoE~\citep{gao2026eegmoe} introduces shared and specific expert groups to decouple domain-specific representations during self-supervised pretraining. Shared structure in MoE can also be configured through RSA-based task affinities ~\citep{vandenhende2020branched}, learned task-expert specialization ~\citep{chen2023modsquad}, compute-constrained architecture search ~\citep{jawahar2023automoe}, or GLU activation statistics in dense language models ~\citep{zhao2026expertweaver}.
SVCCA~\citep{raghu2017svcca}, PWCCA~\citep{morcos2018insights}, and CKA~\citep{kornblith2019similarity} compare neural representations across layers, models, and training stages, while studies in language, vision, and multimodal models show that different depths can encode different information~\citep{tenney2019bert,mischler2024contextual,zhang2025cross}. PRiSE-EEG uses cross-paradigm CKA as an architureture calibration signal to set layer-wise shared-expert counts before training. This makes representation similarity a structural prior rather than only a post-hoc diagnostic.

\subsection{Weak Sensor-Space Priors for EEG}
EEG sensor-space signals do not directly provide source-level functional networks. Nevertheless, electrophysiological recordings still exhibit network-level organization~\citep{brookes2011investigating,hipp2012large,xavier2025consistency} and rapid short-window interactions relevant to dynamic brain states~\citep{vicente2008dynamical,michel2018microstates,van2010microstate,mehra2025zero}. PRiSE-EEG therefore treats static region-network groupings and short-window channel interactions as weak inductive biases rather than hard anatomical assignments. To avoid making these priors overly rigid, we add learnable bias terms that can relax or adjust the predefined sensor-group associations during training. This design preserves coarse spatial guidance while allowing data-driven correction for task-, dataset-, and montage-specific deviations.

\begin{figure*}[t]
\centering
\includegraphics[width=\linewidth]{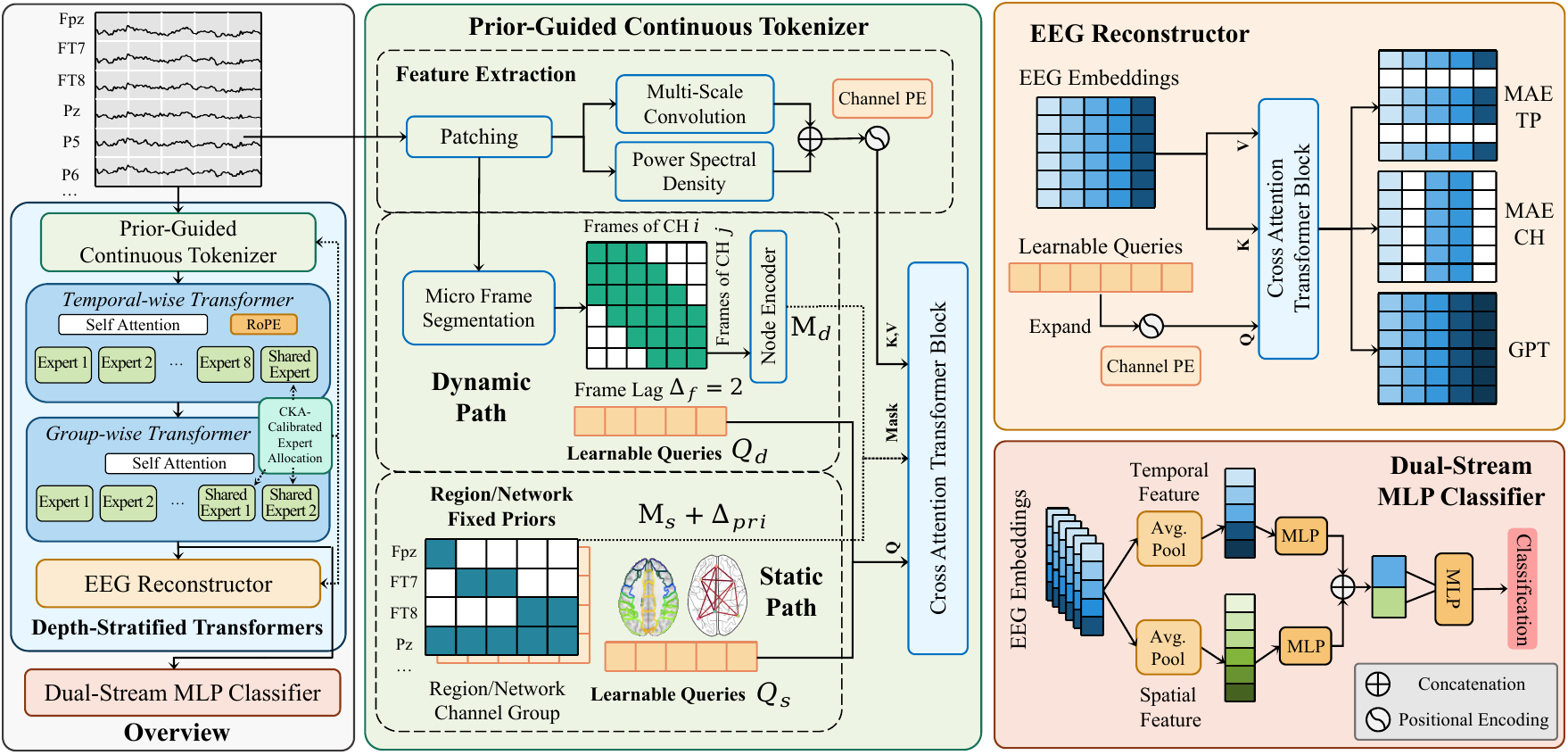}
\caption{The architecture of PRiSE-EEG framework. The Prior-Guided Continuous Tokenizer exploits weak sensor-space prior to extract EEG features and CKA-Calibrated Depth-Stratified Transformers refine expert allocations. The EEG Reconstructor generates the original EEG signal with three pre-train tasks~(MAE-TP, MAE-CH, and GPT), while Dual-Stream MLP Classifiers are applied on both spatial and temporal demension for downstream classification.}
\label{fig:arch}
\end{figure*}

\section{Methods}\label{sec:methods}

\subsection{Overview}

PRiSE-EEG consists of a prior-guided continuous patch tokenizer and a CKA-calibrated depth-stratified expert transformer. Given an EEG batch $\mathbf{x}_p \in \mathbb{R}^{B\times T\times C\times P}$, where $B$ is the batch size, $T$ is the number of 1-second patches, $C$ is the number of channels, and $P$ is the number of sample points, the tokenizer maps $\mathbf{x}_p$ into continuous group representations $\mathbf{h}_0 \in \mathbb{R}^{B\times T\times G\times D_m}$. Here $G$ is the number of static and dynamic groups, and $D_m$ is the model dimension. PRiSE-EEG applies depth-stratified MoE blocks across the tokenizer, temporal/group encoder, and pretraining decoder, with shared-expert ratios determined by block-wise CKA sharedness from a dense pilot model.
Unless otherwise specified, we report results of PRiSE-EEG-B. \textbf{More details are in Appx.~\ref{appx:implementation} and~\ref{appx:model_config}}.

\subsection{Prior-Guided Continuous Tokenization}

We first embed each EEG patch into channel-level representations. Let $\mathbf{x}\in\mathbb{R}^{B\times T\times C\times D_m}$ denote the embedded patch features after temporal-spectral feature extraction and channel positional encoding. 

\begin{equation}
\mathbf{x}=\mathrm{Embed}(\mathbf{x}_p)+\mathbf{E}_c,
\end{equation}

where $\mathbf{E}_c$ is learnable positional encoding. The implementation of feature extraction, channel standardization, and positional encoding is described as Eq.~\ref{eq:multi_scale_embed} and~\ref{eq:embed_concat} in Appendix~\ref{appx:feature_extraction}.

To inject static sensor-space priors, we define coarse region-network groups $\mathcal{P}$. Let $Q_s\in\mathbb{R}^{|\mathcal{P}|\times D_m}$ be learnable static group queries and $\mathbf{M}_s\in\mathbb{R}^{|\mathcal{P}|\times C}$ the corresponding static attention-bias matrix. Since these predefined groups are weak priors rather than hard anatomical assignments, we introduce a learnable relaxation term $\mathbf{\Delta}_{\mathrm{pri}}\in\mathbb{R}^{|\mathcal{P}|\times C}$. Static group attention is computed as
\begin{equation}
\mathrm{Attn}(Q_s,K,V)
=
\mathrm{Softmax}
\left(
\frac{Q_sK^{\top}}{\sqrt{d_k}}
+
\mathbf{M}_s
+
\mathbf{\Delta}_{\mathrm{pri}}
\right)V,
\end{equation}
where $K,V\in\mathbb{R}^{(BT)\times C\times D_m}$ are projected from $\mathbf{x}$, and $d_k$ is the attention head dimension. Static groups provide stable coarse structure, while dynamic groups capture sample-dependent micro-frame interactions. We denote the dynamic bias construction as $\mathbf{M}_d=\mathrm{DynBias}(\mathbf{x}_p;\Delta_f,k_d)$, where $\Delta_f$ is the maximum micro-frame lag and $k_d$ is the number of selected dynamic groups. In practice, $\mathrm{DynBias}(\cdot)$ subdivides each patch into micro-frames, computes lag-limited channel affinities, and selects informative frame-pair groups. Let $Q_d\in\mathbb{R}^{k_d\times D_m}$ be the dynamic group queries. The final query and bias sets are $\mathbf{Q}=[Q_s;Q_d]$ and $\mathbf{M}=[\mathbf{M}_s+\mathbf{\Delta}_{\mathrm{pri}};\mathbf{M}_d]$. The tokenizer output is
\begin{equation}
\mathbf{h}_0
=
\mathrm{CrossAttn}(\mathbf{Q},\mathbf{x};\mathbf{M})
\in
\mathbb{R}^{B\times T\times G\times D_m},
\end{equation}
where $G=|\mathcal{P}|+k_d$. Thus, PRiSE-EEG forms continuous multi-channel group tokens that handle montage heterogeneity while injecting priors. Details are described in Appx.~\ref{appx:static_prior} and~\ref{appx:dynamic_prior}.

\subsection{CKA-Calibrated Depth-Stratified Expert MoE Transformer}
The CKA-calibrated design is applied to all MoE Transformer layers, where each feed-forward block is replaced by a depth-stratified expert layer. Let $\mathbf{h}_l\in\mathbb{R}^{B\times T\times G\times D_m}$ to be input of $l$-th layer.
\\\textbf{Interleaved temporal and group layers.}
The encoder uses $L_E$ alternating Transformer layers: $L_E/2$ temporal-wise layers and $L_E/2$ group-wise layers~\citep{bertasius2021timesformer}. Temporal-wise attention models dependencies across the patch dimension $T$ within each group, while group-wise attention models dependencies across the prior-guided group dimension $G$ within each time patch. Each layer, regardless of its attention type, is followed by an independent depth-stratified MoE FFN:
\begin{equation}
    \tilde{\mathbf{h}}_l=\mathrm{Attn}^{a_l}_l(\mathbf{h}_l),\qquad
    \mathbf{h}_{l+1}=\tilde{\mathbf{h}}_l+\mathrm{DSE}_l(\mathrm{LN}(\tilde{\mathbf{h}}_l)),
\end{equation}
where $a_l\in\{\mathrm{temp},\mathrm{group}\}$ specifies the layer type. This factorized design separates temporal and group-wise interactions while coupling them through alternating layers.
\\\textbf{CKA-Calibrated Depth-Stratified Expert Allocation}. Before training PRiSE-EEG, we run a dense pilot version with the same non-MoE layout and compute CKA on paradigm batches at selected checkpoints $\mathcal{T}$. For layer $l$ in pretraining, let $c^{\mathrm{intra}}_{l,t}$ and $c^{\mathrm{inter}}_{l,t}$ denote the within-paradigm and between-paradigm CKA averages at checkpoint $t$. We define the layer sharedness as
\begin{equation}
s_{l,t}
=
\mathrm{clip}
\left(
\frac{c^{\mathrm{inter}}_{l,t}}
{c^{\mathrm{intra}}_{l,t}+\epsilon},
0,1
\right),
\qquad
\bar{s}_l
=
\frac{1}{|\mathcal{T}|}
\sum_{t\in\mathcal{T}}s_{l,t},
\end{equation}
where $\epsilon$ is a small constant. We convert $\bar{s}_l$ into the shared expert proportion by
\begin{equation}
\rho_l^{\mathrm{sh}}
=
\rho_{\min}
+
(\rho_{\max}-\rho_{\min})
\sigma
\left(
\frac{\bar{s}_l-\tau}{T_{\mathrm{cka}}}
\right),
\end{equation}
where $\rho_{\min}$ and $\rho_{\max}$ bound the shared-expert proportion, $\tau$ is the sharedness threshold, and $T_{\mathrm{cka}}$ is the temperature. Each layer has $E_l=9$ experts: one fixed shared expert and eight allocated by CKA. Both $E_l^{\mathrm{sh}}$ and $\rho_l^{\mathrm{sh}}$ include the fixed expert. We set
\begin{equation}
E_l^{\mathrm{sh}}
=
\mathrm{round}(E_l\rho_l^{\mathrm{sh}}),
\qquad
E_l^{\mathrm{sp}}
=
E_l-E_l^{\mathrm{sh}}.
\end{equation}

Let $\mathcal{E}_l^{\mathrm{sh}}$ denote all shared experts and $\mathcal{F}_{l,e}$ denote expert $e$. The expert block is
\begin{equation}
\mathrm{DSE}_l(\mathbf{h}_l)
=
\sum_{e\in\mathcal{E}_l^{\mathrm{sh}}}\mathcal{F}_{l,e}(\mathbf{h}_l)
+\sum_{e\in\mathcal{K}_l(\mathbf{h}_l)}p_{l,e}(\mathbf{h}_l)\mathcal{F}_{l,e}(\mathbf{h}_l).
\label{eq:dse-output}
\end{equation}
All shared experts process every token; $\mathcal{K}_l(\mathbf{h}_l)$ selects the top-2 specialized experts, and $p_{l,e}$ is the corresponding routing weight. More details of MoE layer are in Appx.~\ref{appx:cka_moe}.
The calibrated allocation is fixed during training; specialized routing remains token-dependent.

\subsection{Stage-Specific Decoder}
PRiSE-EEG uses different lightweight output modules for pretraining and downstream adaptation. 
\textbf{Pretraining Stage}. An EEG reconstructor maps the encoder context $\mathbf{h}_{enc}\in\mathbb{R}^{B\times T\times G\times D_m}$ back to the channel space with learnable channel queries and Transformer decoding layers. It reconstructs both waveform and spectral targets in $\mathbf{x}_{\mathrm{ref}}$, supporting masked temporal-patch reconstruction, masked channel reconstruction, and future-patch prediction. 
\\\textbf{Fine-tuning Stage}. We attach task-specific dual-stream MLP heads. One branch averages over group tokens to obtain temporal features, while the other averages over temporal patches to obtain group-level features. The two features are concatenated, normalized, and mapped to task logits. This head preserves both temporal and group-wise structure with lower cost than fully flattened classifiers.

\subsection{Pretraining and Downstream Adaptation}

During pretraining, PRiSE-EEG optimizes 3 self-supervised and 1 expert balancing loss functions:
\begin{equation}
    \mathcal{L}_{pt} = \frac{1}{3} \left( \mathcal{L}_{\text{MAE-TP}} + \mathcal{L}_{\text{MAE-CH}} + \mathcal{L}_{\text{GPT}} \right) + \beta\mathcal{L}_{\text{Aux}}.
    \label{eq:pretrain-loss}
\end{equation}
where $\mathcal{L}_{\mathrm{MAE\text{-}TP}}$ reconstructs masked temporal patches, $\mathcal{L}_{\mathrm{MAE\text{-}CH}}$ reconstructs masked channels, $\mathcal{L}_{\mathrm{GPT}}$ predicts future patches, $\mathcal{L}_{\mathrm{aux}}$ is the expert load-balancing loss, and $\beta$ is its coefficient.

For downstream adaptation, PRiSE-EEG uses unified multi-dataset fine-tuning by default. One shared backbone is optimized across $K$ downstream datasets with task-specific heads:
\begin{equation}
\mathcal{L}_{\mathrm{ft}}
=
\sum_{k=1}^{K}
\mathcal{L}^{k}_{\mathrm{cls}}
+
\beta\mathcal{L}_{\mathrm{aux}},
\end{equation}
where $\mathcal{L}^{k}_{\mathrm{cls}}$ is the weighted cross-entropy loss for dataset $k$, and $\mathcal{L}_{\mathrm{aux}}$ balances expert usage. Detailed adaptation protocols are given in Appendix~\ref{appx:pt_stage} and~\ref{appx:ft_stage}.

\begin{table}[b]
\centering
\caption{Performance comparison on 12 EEG datasets (Part 1): single-task baselines versus multi-task PRiSE-EEG. Best performance are highlighted in bolded. Statistically significant improvements over the second-best model are indicated by *~(two-sample t-test: p < 0.05).}
\label{tab:results-1}
\setlength{\tabcolsep}{3pt}
\resizebox{1.0\textwidth}{!}{
\begin{tabular}{c c c c c c c c c c c c}
\toprule
\textbf{Dataset} & \textbf{Metrics} & \textbf{EEGNet} & \textbf{Conformer} & \textbf{BENDR}   & \textbf{BIOT}    & \textbf{LaBraM}  & \textbf{EEGPT}    & \textbf{CBraMod}  & \textbf{CSBrain}  & \textbf{REVE}     & \textbf{PRiSE-EEG} \\
\midrule 
\multirow{3}{*}{TUSL}  
& B-Acc                       & 44.94$\pm$1.50 & 54.26$\pm$0.62    & 49.60$\pm$2.02  & 63.96$\pm$1.56  & 64.22$\pm$1.26  & 77.54$\pm$1.95   & 70.91$\pm$1.12   & 66.67$\pm$0.21    & 65.00$\pm$0.79    & \textbf{79.82 $\pm$1.95}* \\
& W-F1                        & 43.95$\pm$1.09 & 46.73$\pm$1.58    & 37.80$\pm$2.42  & 61.51$\pm$1.90  & 51.41$\pm$5.59  & 73.63$\pm$2.70   & 67.51$\pm$2.98   & 60.90$\pm$2.34    & 58.70$\pm$1.98    & \textbf{77.34 $\pm$2.30}* \\
& Kappa                       & 23.06$\pm$0.88 & 25.98$\pm$1.85    & 21.52$\pm$3.80  & 44.19$\pm$2.10  & 37.97$\pm$2.61  & 65.12$\pm$3.48   & 52.18$\pm$3.79   & 43.63$\pm$1.08    & 40.50$\pm$1.65    & \textbf{65.45 $\pm$2.31} \\
\midrule
\multirow{3}{*}
{\makecell[c]{Siena\\Scalp}}
& B-Acc                       & 81.47$\pm$3.08 & 83.27$\pm$1.10    & 78.22$\pm$2.07  & 66.55$\pm$0.51  & 66.36$\pm$1.08  & 81.63$\pm$1.53            & 83.96$\pm$0.55   & 74.11$\pm$0.60    & 70.65$\pm$1.52    & \textbf{86.89$\pm$2.31}* \\
& AUROC                       & 89.83$\pm$1.97 & 88.78$\pm$1.15    & 93.40$\pm$1.03  & 70.64$\pm$1.88  & 77.39$\pm$1.80  & \textbf{93.71$\pm$1.54}   & 93.15$\pm$1.49   & 89.07$\pm$2.39    & 83.56$\pm$1.28    & 89.12$\pm$2.30 \\
& AUC-PR                      & 99.78$\pm$0.05 & 99.72$\pm$0.02    & 99.82$\pm$0.03  & 99.05$\pm$0.13  & 99.50$\pm$0.07  & 99.85$\pm$0.03            & 99.84$\pm$0.04   & 99.72$\pm$0.11    & 99.64$\pm$0.03    & \textbf{99.85$\pm$0.03} \\
\midrule
\multirow{3}{*}{SEED} 
& B-Acc                       & 51.40$\pm$0.42 & 55.79$\pm$0.50    & 60.92$\pm$0.53  & 64.57$\pm$1.00  & 64.52$\pm$0.53  & 67.53$\pm$0.29   & 70.86$\pm$1.80            & 71.92$\pm$0.39    & 70.29$\pm$0.78    & \textbf{73.92$\pm$0.31}* \\
& W-F1                        & 51.62$\pm$0.84 & 54.27$\pm$0.41    & 59.98$\pm$0.94  & 64.01$\pm$0.80  & 64.60$\pm$0.35  & 67.77$\pm$0.05   & 70.09$\pm$2.06            & 71.84$\pm$0.20    & 69.62$\pm$0.72    & \textbf{73.29$\pm$0.22}* \\
& Kappa                       & 27.43$\pm$0.79 & 33.09$\pm$0.56    & 41.55$\pm$0.79  & 47.03$\pm$1.50  & 47.10$\pm$0.79  & 51.92$\pm$0.61   & 56.46$\pm$2.70            & 58.02$\pm$0.59    & 55.59$\pm$1.18    & \textbf{61.02$\pm$0.45}* \\
\midrule         
\multirow{3}{*}{SEED-V}       
& B-Acc                       & 29.61$\pm$0.28 & 35.37$\pm$0.23    & 22.49$\pm$0.19  & 30.48$\pm$0.24  & 39.60$\pm$1.09  & 38.33$\pm$0.69   & 40.98$\pm$1.58            & 38.02$\pm$0.23    & 40.84$\pm$0.98    & \textbf{47.40$\pm$0.25}* \\
& W-F1                        & 27.49$\pm$1.20 & 34.87$\pm$0.38    & 22.22$\pm$1.95  & 31.05$\pm$0.30  & 39.55$\pm$0.85  & 36.96$\pm$0.89   & 41.18$\pm$1.46            & 38.07$\pm$1.03    & 40.58$\pm$2.00    & \textbf{48.39$\pm$0.42}* \\
& Kappa                       & 10.06$\pm$0.39 & 17.72$\pm$0.25    & 3.86$\pm$0.03  & 14.07$\pm$0.28  & 24.17$\pm$1.20  & 22.08$\pm$1.06   & 25.58$\pm$1.87            & 22.48$\pm$0.42    & 25.59$\pm$1.48    & \textbf{34.84$\pm$0.45}* \\
\midrule         
\multirow{3}{*}{PhysioMI}     
& B-Acc                             & 57.25$\pm$0.73 & 55.23$\pm$0.19    & 48.01$\pm$0.29  & 61.53$\pm$1.54  & 61.73$\pm$0.89  & 62.92$\pm$0.32   & 61.74$\pm$0.91            & 63.04$\pm$0.74    & 64.80$\pm$0.53   & \textbf{65.55$\pm$0.38}* \\
& W-F1                              & 56.68$\pm$0.97 & 40.58$\pm$0.50    & 46.57$\pm$3.06  & 61.58$\pm$1.97  & 61.77$\pm$1.16  & 62.72$\pm$0.34   & 61.79$\pm$1.69            & 63.08$\pm$0.81    & 64.84$\pm$0.94   & \textbf{65.83$\pm$0.56}* \\
& Kappa                             & 42.91$\pm$0.95 & 46.50$\pm$0.38    & 30.66$\pm$0.39  & 48.75$\pm$2.72  & 52.22$\pm$1.19  & 53.12$\pm$1.63   & 52.22$\pm$1.00            & 53.80$\pm$0.76    & 53.04$\pm$0.85   & \textbf{54.78$\pm$0.69}* \\
\midrule         
\multirow{3}{*}{Mimul-11}     
& B-Acc                       & 45.92$\pm$0.59 & 49.25$\pm$0.81    & 49.68$\pm$0.94  & 41.73$\pm$0.28  & 48.57$\pm$1.04  & 47.09$\pm$0.66   & 50.51$\pm$0.63            & 42.67$\pm$0.77    & 45.00$\pm$0.62    & \textbf{51.33$\pm$0.24}* \\
& W-F1                        & 49.29$\pm$1.15 & 46.51$\pm$1.70    & 58.49$\pm$1.22  & 50.59$\pm$0.56  & 57.95$\pm$0.96  & 56.60$\pm$0.95   & 58.09$\pm$0.61            & 51.93$\pm$0.87    & 54.63$\pm$0.71    & \textbf{58.82$\pm$0.23}* \\
& Kappa                       & 20.28$\pm$1.62 & 21.98$\pm$1.05    & 32.74$\pm$3.52  & 15.96$\pm$1.08  & 31.91$\pm$2.54  & 26.23$\pm$2.89   & \textbf{33.63$\pm$1.80}   & 18.13$\pm$2.05    & 23.10$\pm$1.31    & 31.49$\pm$0.98 \\
\bottomrule
\end{tabular}
}
\vspace{-7pt}
\end{table}

\section{Experiments}
\subsection{Datasets}
\textbf{Pre-training Datasets.}
To learn robust and generalizable representations, PRiSE-EEG is pre-trained on a diverse corpus combining 17 public datasets, totaling over 28,000 hours of EEG signals.
Detailed information about pre-training datasets is in Appx.~\ref{appx:data}.
\\\textbf{Fine-tuning Datasets.}
We conduct experiments across \textbf{12 EEG datasets} spanning \textbf{10 task categories}, including: 
(1) Seizure Detection: Siena~\citep{detti2020siena}; (2) Emotion Recognition: SEED~\citep{zheng2015seed}, SEED-V~\citep{liu2022seed-v}; (3) Motor Imagery: PhysioMI~\citep{Schalk2004physiomi}, Mimul-11~\citep{jeong2020mimul-11}; (4) Mental Stress: Workload~\citep{zyma2019workload}; (5) Sleep Staging: HMC~\citep{alvarez2021hmc}; (6) Anomalous Event Detection: TUEV~\citep{harati2015tuev}; (7) Abnormal Classification: TUAB~\citep{lopez2015tuab}; (8) Visual Target Detection: Things-EEG-2~\citep{gifford2022things-eeg-2}; (9) Alzheimer’s Disease Identification: ADFTD~\citep{miltiadous2023adftd}; and (10) Slowing Event Classification: TUSL~\citep{weltin2017tusl} (details in Appx.~\ref{appx:data}).

\subsection{Experimental Setup}
\textbf{Data Preprocessing.} 
We utilize an MNE-based~\citep{Gramfort2013mne} pipeline to resample, filter, and normalize EEG signals, facilitating reproducible and scalable feature extraction.
\\\textbf{Evaluation Metrics.} 
To account for class imbalance in downstream tasks, we follow the protocol in \citet{jiang2024neurolm} and employ Balanced Accuracy, Cohen’s Kappa, and Weighted F1 for multi-class classification, and Balanced Accuracy, AUC-PR, and AUROC for binary classification. 
Model optimization dynamics and consistency are assessed using cosine similarity and expert router statistics.
\\\textbf{Baselines.}
We compare PRiSE-EEG against two supervised benchmarks (EEGNet~\citep{lawhern2018eegnet}, EEGConformer~\citep{song2022eegconformer}) and seven EEG foundation models (BENDR~\citep{kostas2021bendr}, BIOT~\citep{yang2023biot}, LaBraM~\citep{jiang2024large}, REVE~\citep{Ouahidi2025reve}, CBraMod~\citep{wang2024cbramod}, EEGPT~\citep{wang2024eegpt}, and CSBrain~\citep{zhou2025csbrain}).
\textbf{Experimental details are in Appx.~\ref{appx:exp_setting}}.

\begin{table}[t]
\centering
\caption{Performance comparison on 12 EEG datasets (Part 2): single-task baselines versus multi-task PRiSE-EEG. Best performance are highlighted in bolded. Statistically significant improvements over the second-best model are indicated by *~(two-sample t-test: p < 0.05).}
\label{tab:results-2}
\setlength{\tabcolsep}{3pt}
\resizebox{1.0\textwidth}{!}{
\begin{tabular}{c c c c c c c c c c c c}
\toprule
\textbf{Dataset} & \textbf{Metrics} & \textbf{EEGNet} & \textbf{Conformer}  & \textbf{BENDR}    & \textbf{BIOT}     & \textbf{LaBraM}   & \textbf{EEGPT}    & \textbf{CBraMod}  & \textbf{CSBrain}  & \textbf{REVE}     & \textbf{PRiSE-EEG} \\
\midrule
\multirow{3}{*}{Workload}
& B-Acc                             & 66.74$\pm$1.92 & 67.47$\pm$0.30     & 62.67$\pm$0.99   & 63.67$\pm$0.59   & 57.93$\pm$0.86   & 69.31$\pm$1.69   & 74.23$\pm$1.36   & 71.27$\pm$0.75    & 69.53$\pm$2.65    & \textbf{82.47$\pm$0.30}* \\
& AUROC                             & 77.48$\pm$1.26 & 76.94$\pm$0.42     & 71.38$\pm$3.02   & 75.09$\pm$1.42   & 78.29$\pm$0.48   & 74.70$\pm$0.98   & 82.53$\pm$0.75   & 79.63$\pm$1.20    & 77.07$\pm$4.26    & \textbf{89.21$\pm$0.19}* \\
& AUC-PR                            & 49.98$\pm$1.50 & 50.30$\pm$1.14     & 50.24$\pm$3.75   & 69.09$\pm$2.25   & 60.15$\pm$0.55   & 55.93$\pm$1.92   & 61.48$\pm$1.90   & 55.10$\pm$2.21    & 64.67$\pm$3.20    & \textbf{70.75$\pm$0.47}* \\
\midrule
\multirow{3}{*}{TUEV} 
& B-Acc                             & 58.15$\pm$2.31 & 58.00$\pm$0.63     & 67.81$\pm$0.61   & 59.43$\pm$0.70   & 65.85$\pm$0.77   & 61.31$\pm$1.36   & 67.64$\pm$0.47   & 71.20$\pm$1.34    & 63.24$\pm$0.73    & \textbf{75.28$\pm$0.51}* \\
& W-F1                              & 74.77$\pm$1.39 & 73.02$\pm$0.25     & 81.19$\pm$1.54   & 77.42$\pm$2.03   & 78.22$\pm$2.36   & 80.00$\pm$0.91   & 81.31$\pm$0.54   & 85.07$\pm$0.85    & 84.97$\pm$0.33    & \textbf{88.25$\pm$0.51}* \\
& Kappa                             & 59.55$\pm$2.91 & 56.69$\pm$0.21     & 66.04$\pm$2.78   & 63.12$\pm$3.05   & 64.49$\pm$3.54   & 66.87$\pm$1.35   & 67.22$\pm$1.02   & 74.67$\pm$1.44    & 75.12$\pm$0.76    & \textbf{80.90$\pm$0.62}* \\
\midrule
\multirow{3}{*}{TUAB} 
& B-Acc                             & 79.87$\pm$0.46 & 81.78$\pm$0.53     & 81.67$\pm$0.70   & 81.23$\pm$0.42   & 80.74$\pm$0.43   & 81.75$\pm$0.65   & 80.89$\pm$0.36   & 78.97$\pm$0.17    & 80.53$\pm$0.12    & \textbf{82.22$\pm$0.12}* \\
& AUROC                             & 88.78$\pm$0.25 & 89.58$\pm$0.49     & 88.08$\pm$0.45   & 88.36$\pm$0.27   & 86.72$\pm$1.48   & 89.60$\pm$1.05   & 87.81$\pm$0.45   & 85.33$\pm$1.51    & 87.43$\pm$1.04    & \textbf{89.93$\pm$0.31} \\
& AUC-PR                            & 87.86$\pm$0.67 & 90.60$\pm$0.53     & 88.18$\pm$1.41   & 88.91$\pm$0.33   & 87.22$\pm$1.49   & 90.00$\pm$1.56   & 88.51$\pm$0.53   & 82.73$\pm$4.69    & 86.00$\pm$2.20    & \textbf{90.11$\pm$0.15} \\
\midrule
\multirow{3}{*}
{\makecell[c]{Things-\\EEG-2}} 
& B-Acc                             & 51.27$\pm$0.29 & 60.30$\pm$0.50     & 63.76$\pm$1.57   & 51.87$\pm$0.09   & 51.48$\pm$0.21   & 60.83$\pm$0.82   & 60.25$\pm$1.19   & 57.20$\pm$0.29    & 61.03$\pm$0.90    & \textbf{65.15$\pm$0.48}* \\
& AUROC                             & 62.10$\pm$0.10 & 68.42$\pm$1.39     & 77.54$\pm$0.70   & 66.17$\pm$0.73   & 52.78$\pm$0.23   & 73.53$\pm$1.56   & 75.11$\pm$1.92   & 62.67$\pm$0.53    & 66.77$\pm$0.52    & \textbf{86.63$\pm$0.35}* \\
& AUC-PR                            & 16.82$\pm$0.22 & 22.31$\pm$1.37     & 40.89$\pm$1.20   & 18.47$\pm$0.53   & 12.39$\pm$0.05   & 31.91$\pm$3.21   & 34.19$\pm$1.92   & 21.53$\pm$0.42    & 22.87$\pm$1.72    & \textbf{42.82$\pm$1.19}*  \\
\midrule
\multirow{3}{*}{HMC} 
& B-Acc                             & 58.31$\pm$0.95 & 66.68$\pm$0.94     & 70.25$\pm$1.00   & 71.33$\pm$0.50   & 70.68$\pm$0.89   & 71.93$\pm$0.41   & 71.48$\pm$0.40   & 72.18$\pm$0.24    & 71.13$\pm$0.26    & \textbf{74.33$\pm$0.21}* \\
& W-F1                              & 58.37$\pm$1.87 & 64.73$\pm$1.35     & 72.30$\pm$1.35   & 72.62$\pm$0.33   & 71.73$\pm$1.50   & 72.79$\pm$0.64   & 72.76$\pm$0.55   & 74.16$\pm$0.27    & 73.10$\pm$0.22    & \textbf{77.31$\pm$0.20}* \\
& Kappa                             & 48.29$\pm$1.94 & 57.15$\pm$1.05     & 64.51$\pm$1.50   & 64.94$\pm$0.36   & 64.04$\pm$1.30   & 64.97$\pm$0.53   & 64.86$\pm$0.41   & 66.43$\pm$0.16    & 65.20$\pm$0.29    & \textbf{70.69$\pm$0.19}* \\
\midrule
\multirow{3}{*}{ADFTD} 
& B-Acc                             & 37.07$\pm$1.24 & 36.68$\pm$0.45     & 41.83$\pm$0.37   & 42.31$\pm$1.25   & 37.86$\pm$2.07   & 50.81$\pm$2.28   & 52.20$\pm$0.93   & 48.96$\pm$2.65    & 44.38$\pm$0.75    & \textbf{58.75$\pm$0.66}* \\
& W-F1                              & 30.65$\pm$6.15 & 33.09$\pm$1.46     & 43.10$\pm$0.51   & 44.01$\pm$2.37   & 35.31$\pm$2.91   & 50.10$\pm$1.52   & 55.29$\pm$1.50   & 50.17$\pm$3.76    & 50.55$\pm$0.43    & \textbf{64.12$\pm$0.63}* \\
& Kappa                             & 17.14$\pm$15.40 & 5.43$\pm$0.83     & 16.35$\pm$0.68   & 16.78$\pm$2.11   & 6.08$\pm$2.92   & 25.68$\pm$0.99   & 31.38$\pm$1.82   & 24.96$\pm$4.13    & 23.17$\pm$0.83    & \textbf{39.41$\pm$0.79}* \\
\bottomrule
\end{tabular}
}
\end{table}

\subsection{Main Results}
Tab.~\ref{tab:results-1} and~\ref{tab:results-2} report the results of \textbf{PRiSE-EEG} (\textbf{basic version: 41.06M parameters with 18.05M activated}) across diverse benchmarks.
Baselines are trained or fine-tuned separately per task, whereas PRiSE-EEG uses unified multi-task fine-tuning.
PRiSE-EEG results are mean $\pm$ standard deviation over 10 runs. Best results are in \textbf{bold}, and statistically significant improvements over the second-best model are marked by *.
\textbf{PRiSE-EEG} achieves strong performance on most tasks.
It shows large gains on disease-related and cognitive tasks, including ADFTD, SEED-V, and Workload, improving primary metrics by \textbf{6-8\%} over existing EEG-FMs.
Consistent \textbf{2-5\%} gains are observed on TUEV, TUSL, PhysioMI, SEED, and HMC.
For tasks such as TUAB and Mimul-11, where the gain in B-Acc is approximately 1\%, \textbf{PRiSE-EEG} displays lower variance with enhanced stability.
On Things-EEG-2, it achieves a \textbf{>9\%} AUROC gain, indicating stronger fine-grained visual target discrimination.

Compared with supervised baselines (\textit{e.g.}, EEGNet and Conformer), \textbf{PRiSE-EEG} benefits from self-supervised pretraining on large EEG corpora, reducing overfitting to task-specific labeled data.
It also outperforms existing EEG-FMs, which we attribute to prior-guided continuous tokenization and the CKA-calibrated depth-stratified expert transformer.
\textbf{Under single-task fine-tuning, the pretrained model surpasses existing EEG-FMs on most tasks} (Tab.~\ref{tab:detailed-config-result}), showing that the gains do not rely solely on unified multi-dataset fine-tuning.
Overall, these results support the effectiveness of combining weak sensor-space priors with CKA-calibrated depth-aware expert allocation for robust cross-paradigm EEG representation learning.
\textbf{Additional results on fine-tuning strategies, scaling, computational costs and representation-analysis are in Appx.~\ref{appx:extra_result} and~\ref{appx:scaling}, \ref{appx:compute}}.

\subsection{Ablation Study}
\begin{figure*}[t]
\centering
\includegraphics[width=\textwidth]{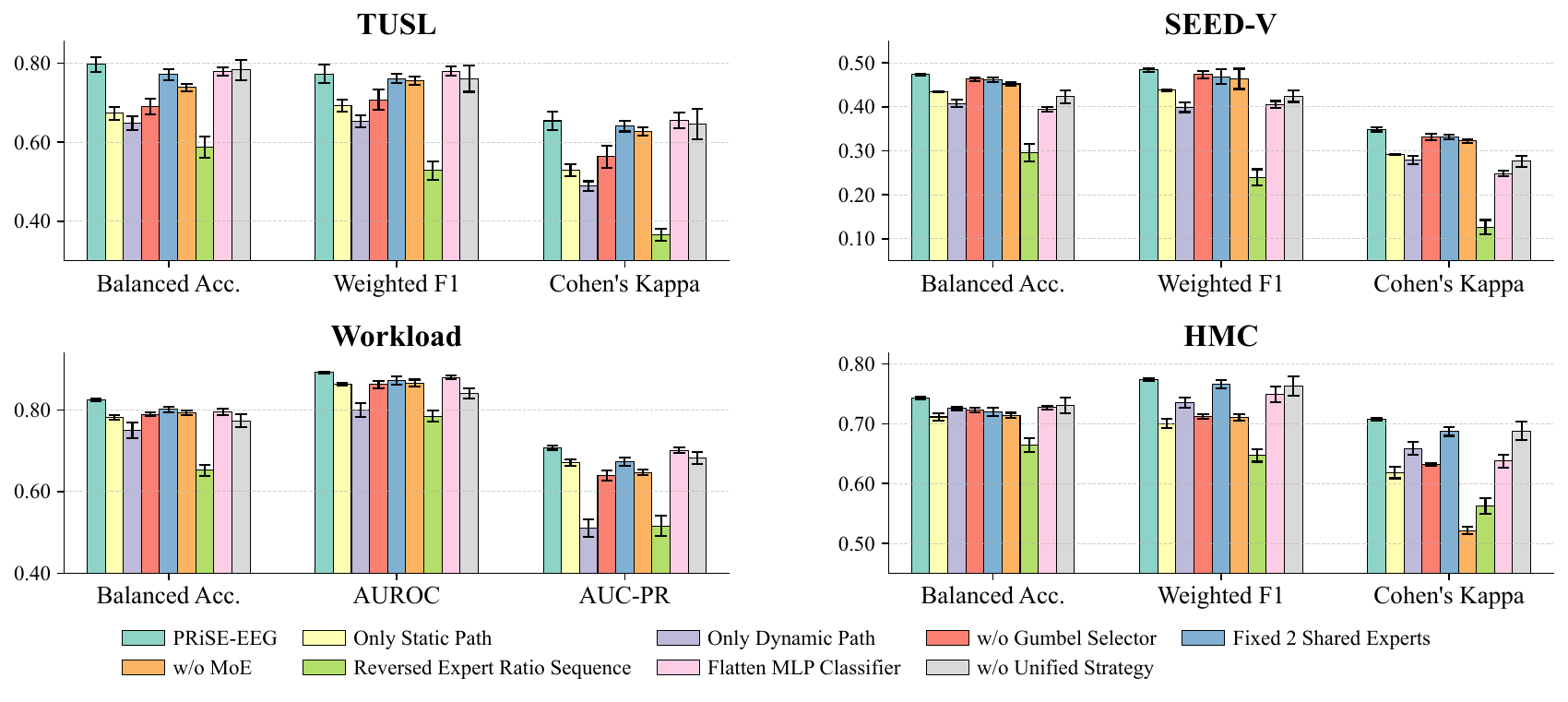}
\caption{Ablations of tokenization, expert allocation, classifier, and unified fine-tuning. ``Fixed 2 shared experts'' selects two shared experts from the expert pool, giving three shared experts in total.}
\label{fig:ablation}
\end{figure*}

\begin{figure}[b]
\centering
\includegraphics[width=\linewidth]{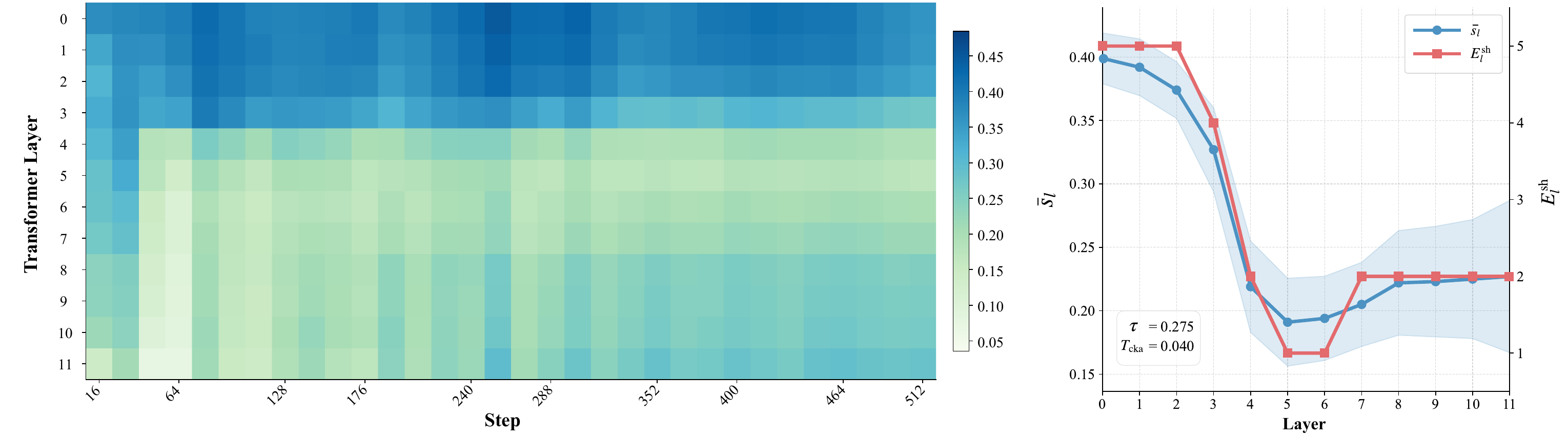}
\caption{
Layer-wise CKA sharedness over steps 16-512 (left) and total shared-expert counts (right), including the fixed expert; $\tau=0.275$, $T_{\mathrm{cka}}=0.040$.
}
\label{fig:cka_allocation}
\end{figure}

Fig.~\ref{fig:ablation} reports ablations on four representative datasets, covering prior-guided tokenization, CKA-calibrated depth-stratified experts, the classification head and unified fine-tuning.

\textbf{Prior-guided continuous tokenization.}
Combining static and dynamic paths achieves higher Balanced Accuracy than either path alone on all four tasks, showing that both coarse sensor-space priors and sample-dependent short-window interactions are useful for EEG patch formation. The static path is more effective on TUSL, SEED-V, and Workload, whereas the dynamic path performs better on HMC. Removing Gumbel selection lowers Balanced Accuracy by 1.09-10.71\%, with the largest drop on TUSL.

\textbf{CKA-calibrated depth-stratified experts.}
Replacing MoE with the standard dense FFN lowers Balanced Accuracy by 2.21-5.94\%. Depth-stratified allocation improves over the fixed-sharing baseline by 1.28-2.70\%, while reversing the allocation order causes larger losses of 7.93-21.02\%. These results support both the tested expert structure and its depth-dependent allocation rather than uniform sharing. Detailed results and computational costs are provided in Appx.~\ref{appx:extra_ablation} and~\ref{appx:compute}.

\textbf{Classifier and fine-tuning strategy.}
The dual-stream head achieves comparable or stronger performance than the flattened MLP while reducing head parameters by over 90\%, suggesting that separate temporal and group pooling preserves useful structure without parameter explosion.
The drop without unified fine-tuning further supports joint adaptation within this configuration, without establishing which component accounts for most of the overall gain.

\subsection{Representation and Routing Analysis}
\begin{figure}[t]
\centering
\captionsetup{skip=5pt}
\includegraphics[width=0.695\linewidth]{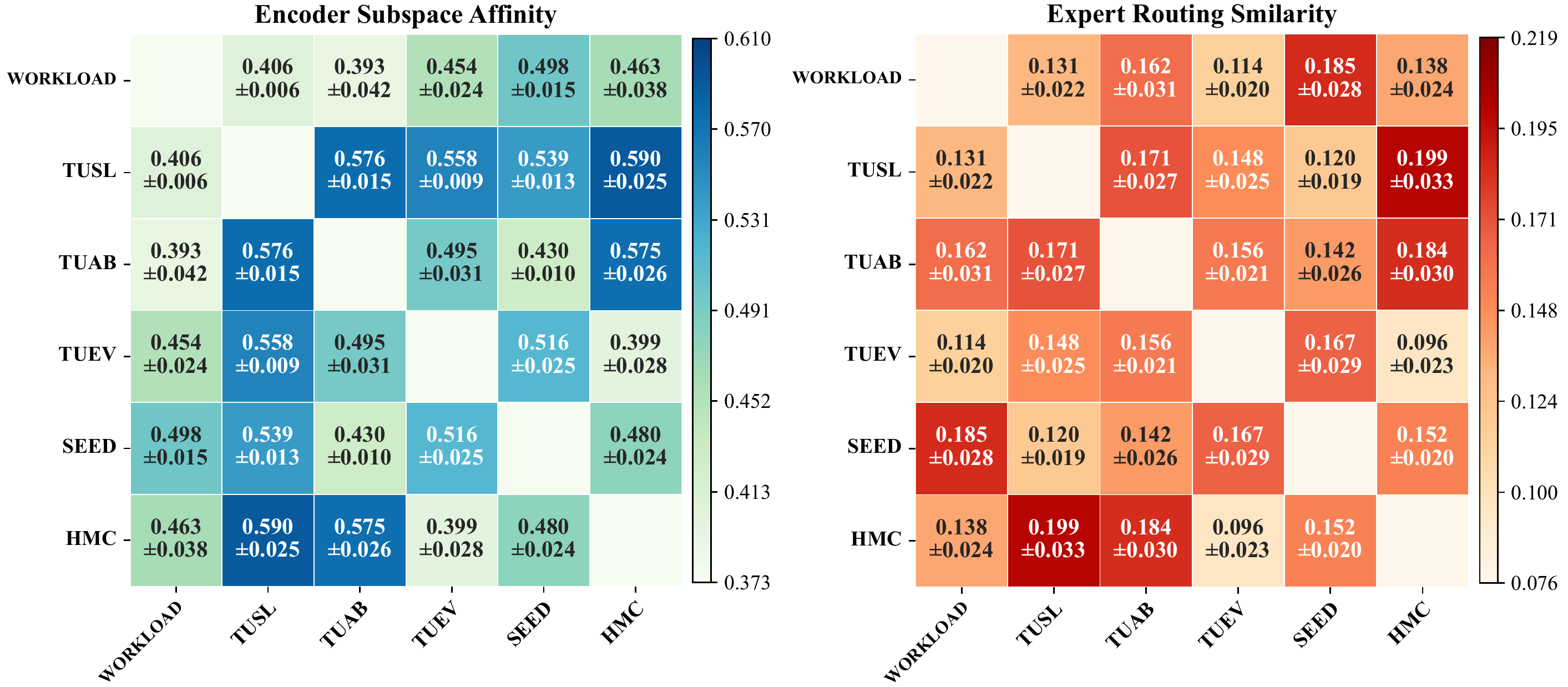}
\caption{Gradient subspace affinity (rank=5) and similarity of routed expert distribution among different datasets.}
\label{fig:sim}
\end{figure}

We analyze PRiSE-EEG from three perspectives: prior-group attribution, CKA-calibrated expert allocation, and the relation between gradient geometry and expert routing.
\\\textbf{Attribution of prior groups.}
We treat static and dynamic attention biases in the tokenizer as attribution targets and compute Integrated Gradients~\citep{sundararajan2017axiomatic} of the target logits with respect to each group bias using a zero baseline.
Absolute attributions are accumulated across heads and temporal positions to identify influential prior groups.
As shown in Tab.~\ref{tab:top3_group}, PRiSE-EEG consistently assigns high attribution to frontal and temporal regions for emotion-related tasks, while occipital and parietal groups become more prominent in visual or decision-related tasks.
Dynamic groups further reveal lag-dependent shifts of attended regions, suggesting that the tokenizer uses both stable spatial priors and sample-dependent short-window interactions.
Additional Grad-CAM~\citep{selvaraju2017grad} visualizations are provided in Appx.~\ref{appx:visual}.
\\\textbf{CKA-calibrated shared expert allocation.}
Fig.~\ref{fig:cka_allocation} visualizes the layer-wise CKA sharedness used to configure the depth-stratified expert transformer.
The CKA map shows a clear depth-wise transition: shallow layers maintain higher cross-paradigm alignment, whereas middle and deeper layers become more specialized.
Accordingly, the sigmoid mapping assigns more shared experts to early layers and more specialized capacity to later layers.
CKA guides layer-wise sharing; fixed-count and reversed-order ablations assess the value of depth-dependent allocation.
\\\textbf{Gradient subspace and expert routing.}
Fig.~\ref{fig:sim} compares gradient subspace affinity with expert routing similarity across datasets.
Encoder gradients are collected for 512 optimization steps under dataset-specific heads, projected to a lower-dimensional space, and summarized by rank-$k$ PCA bases.
For each dataset pair, subspace affinity measures the overlap between dominant update directions, while routing similarity is computed from cumulative expert usage.
Datasets with more similar gradient subspaces tend to induce more similar routing patterns, such as SEED-TUEV and HMC-TUAB.
This suggests that the depth-stratified experts do not route samples arbitrarily, but reflect dataset-level optimization geometry. Detailed computation method is described in Appx.~\ref{appx:gradient_analysis}.

\begin{table}[t]
\centering
\small
\captionsetup{skip=5pt}
\caption{The significant attribution groups for static and dynamic attention biases via Integrated Gradient.}
\label{tab:top3_group}
\setlength{\tabcolsep}{2pt}
\resizebox{0.5\linewidth}{!}{
\begin{tabular}{c c c c c}
\toprule
\multirow{2}{*}{\textbf{Path}}
& \multicolumn{2}{c}{\textbf{SEED}}
& \multicolumn{2}{c}{\textbf{THINGS-EEG-2}} \\
\cmidrule(lr){2-3}\cmidrule(lr){4-5}
& \textbf{Group Name} & \textbf{Focus Region}
& \textbf{Group Name} & \textbf{Focus Region} \\
\midrule
\multirow{3}{*}{\makecell[c]{\textbf{Static}}}
& Frontal           & -  & Occipital        & - \\
& \makecell[c]{Right Temporal}     & -  & \makecell[c]{Visual Network}   & - \\
& \makecell[c]{Salience Network}  & -  & \makecell[c]{Dorsal Attention} & - \\
\midrule
\multirow{3}{*}[-3.5ex]{\textbf{Dynamic}}
& $\mathrm{Lag}$-0   & \makecell[c]{Frontal\\Temporal}          & $\mathrm{Lag}$-0 & Occipital \\
\cmidrule(lr){2-5}
& $\mathrm{Lag}$-1   & \makecell[c]{Frontal\\Temporal} & $\mathrm{Lag}$-1 & \makecell[c]{Occipital\\Parietal} \\
\cmidrule(lr){2-5}
& $\mathrm{Lag}$-2   & \makecell[c]{Frontal\\Temporal\\Parietal}  & $\mathrm{Lag}$-2 & \makecell[c]{Temporal\\Frontal} \\
\bottomrule
\end{tabular}
}
\end{table}

\section{Conclusion}
This work presents \textbf{PRiSE-EEG}, a prior-guided EEG foundation model for cross-paradigm representation learning.
PRiSE-EEG forms continuous multi-channel patches with weak sensor-space priors and short-window interactions, and allocates shared and specialized expert capacity across layers using CKA-calibrated sharedness.
Evaluated on 12 public benchmarks spanning 10 EEG task categories, PRiSE-EEG achieves strong performance in evaluations across multiple EEG paradigms.
Further ablations and analyses show that prior-guided tokenization, depth-stratified expert allocation, and unified fine-tuning jointly improve cross-paradigm adaptation.
Overall, PRiSE-EEG provides a compact framework for balancing shared EEG regularities with task-specific specialization.


\subsection*{AI use statement}
In this work, generative AI tools were used primarily to improve the clarity, grammar, and readability of the manuscript. They were not used for research ideation, methodology design, implementation, data processing, experimentation, or the analysis and interpretation of results. All AI-assisted edits were reviewed and verified by the authors, who take full responsibility for the final manuscript.

\subsection*{Ethics Statement}
This study uses existing EEG datasets made available for research, with no new data collection or participant recruitment. We follow the datasets' access conditions and ethical protocols, do not attempt participant re-identification, and report no participant-level information. EEG recordings contain sensitive physiological information, and differences in populations and acquisition settings may introduce dataset-specific biases. Our results do not establish clinical validity across populations. PRiSE-EEG is intended for research use; deployment in clinical or other high-stakes settings requires independent validation, appropriate data governance, and qualified professional oversight.

\subsection*{Reproducibility Statement}
Code for reproducing the experiments is available at \url{https://github.com/xw1216/PRiSE-EEG}. Section~\ref{sec:methods} describes the architecture, expert allocation, and training objectives. Appendices~\ref{appx:data}, \ref{appx:implementation}, and~\ref{appx:exp_setting} detail the datasets, implementation, preprocessing, data splits, training settings, and evaluation protocols. Appendix~\ref{appx:gradient_analysis} describes the gradient and routing analyses, and Appendix~\ref{appx:extra_result} reports additional results and ablations.

\bibliography{sample-base}
\bibliographystyle{iclr2027_conference}


\newpage
\appendix

This supplementary material presents additional details, analyses, and experiments to support our primary findings. 
The document is structured as follows:
\begin{itemize}
    \item Appendix~\ref{appx:data} describes the datasets utilized for pre-training and fine-tuning.
    \item Appendix~\ref{appx:implementation} elaborates on the specific implementation details of the model.
    \item Appendix~\ref{appx:gradient_analysis} details the method to analyze the gradient and routing dynamics.
    \item Appendix~\ref{appx:exp_setting} details the complete experimental configurations.
    \item Appendix~\ref{appx:extra_result} presents supplementary experimental results and ablation studies evaluating the impact of different configurations.
    \item Appendix~\ref{appx:visual} contains additional visualizations of classification results and saliency topomaps to further illustrate the model's interpretability.
    \item Appendix~\ref{appx:scaling} provides an empirical investigation into the scaling behavior governing EEG representations.
    \item Appendix~\ref{appx:limit} concludes with a discussion of current limitations and outlines potential avenues for future research.
\end{itemize}

\section{Datasets Description}\label{appx:data}
\subsection{Pre-training Datasets}
The pre-training corpus for \textbf{PRiSE-EEG} comprises \textbf{17} diverse, publicly accessible EEG datasets.
This multi-site corpus spans various experimental protocols, including clinical monitoring (\textit{e.g.}, HBN, TUEG, Siena), cognitive assessments (\textit{e.g.}, SEED series, Motor Imagery), and brain-computer interface (BCI) tasks (\textit{e.g.}, BCIC, P300).
Comprehensive specifications for each dataset, including specific citations, recording durations, and functional descriptions, are summarized in Tab.~\ref{tab:pretrain-dataset}.
To prevent pretraining data leakage, all TUEG recordings whose subject identifiers overlap with TUAB, TUEV, or TUSL evaluation subjects are removed from the pre-training corpus.

\begin{table}[!ht]
    \centering
    \setlength{\tabcolsep}{1mm}
    \caption{Detailed information about pre-training datasets. Window durations are in seconds.}
    \label{tab:pretrain-dataset}
    \resizebox{0.95\textwidth}{!}{
    \begin{tabular}{ccccccm{6.5cm}}
    \toprule
    Dataset &  Category & \#Channel & Duration & \#Train & \#Valid & \makecell[c]{Description}\\
    \midrule
    TUEG~\citep{obeid2016temple} & \makecell{Clinical\\Recordings}  &  19-22 & 60 & 1515391 & 75436 & A rich archive of clinical EEG recordings collected at Temple University Hospital. \\ 
    \midrule
    HBN~\citep{shirazi2024hbn} & \makecell{Multiple\\Categories} & 70 & 15 & 395605 & 42376 & EEG data and behavioral responses collected during EEG experiments from more than 3000 participants (5-21) performing six distinct tasks in the HBN project. \\
    \midrule
    SEED-IV~\citep{8283814} & \makecell{Emotion\\Recognition}& 60 & 10 & 13678 & 977 & A emotion EEG dataset that 15 subjects watch 72 film clips which will induce happiness, sadness, fear or neutral emotions.\\
    \midrule
    SEED-GER~\citep{liu2022identifying} & \makecell{Emotion\\Recognition} & 60 & 10 & 8440 & 919 & A emotion dataset eight German subjects watch 20 film clips (positive, neutral, negative) as stimuli. \\
    \midrule
    SEED-FRA~\citep{liu2022identifying}  & \makecell{Emotion\\Recognition} & 60 & 10 & 6846 & 978 & Eight French subjects watch 21 film clips in French (positive, neutral, negative) as stimuli. \\
    \midrule
    BCIC-1A~\citep{BLANKERTZ2007539} & \makecell{Motor\\Imagery} & 43 & 8 & 3155 & 535 & EEG recordings for motor imagery tasks, where subjects imagined moving either their left hand, right hand, or foot. \\
    \midrule
    Emobrain~\citep{savran2006emobrain} & \makecell{Emotion\\Recognition} & 54 & 10 & 1370 & 405 & Multimodal emotion detection dataset using brain signals (EEG, fNIRS) from 5 male subjects. \\
    \midrule
    \makecell{Grasp and Lift\\\citep{Luciw2014}} & \makecell{Motor\\Execution} & 32 & 5 & 7003 & 1390 & Grasp and lift action from 12 subjects in total,  10 series of trials for each subject.\\
    \midrule
    \makecell{Inria BCI P300\\\citep{Margaux2012}} & ERP & 56 & 5 & 13647 & 7901 & A P300-based spelling dataset including 26 subjects. \\
    \midrule
    \makecell{Resting State\\\citep{10.3389/fnins.2017.00425}} & Resting & 64 & 10 & 981 & 100 & A dataset comprising 22 subjects for a resting eyes closed and eyes open. \\
    \midrule
    \makecell{Raw EEG Data\\\citep{10.3389/fnins.2019.01292}} & \makecell{Visual\\Stimulus} & 64 & 30 & 4806 & 319 & A dataset recorded during reported Information-Integration categorization and reported multidimensional Rule-Based categorization tasks. \\
    \midrule
    \makecell{SPIS Resting State\\\citep{9034192}} & Resting & 64 & 10 & 270 & 30 & A resting-state EEG from 10 subjects contains 2.5 minutes of eyes-open and 2.5 minutes of eyes-closed. \\
    \midrule
    \makecell{Target Versus Non-Target\\\citep{korczowski:hal-02172347}} & ERP & 32 & 15 & 3403 & 333 & Dataset contains EEG recordings of 50 subjects playing to a visual P300 Brain-Computer Interface (BCI) video game. \\
    \midrule
    \makecell{BrainLat~\citep{prado2023brainlat}} & \makecell{Alzheimer's\\Disease}  & 48 & 10 & 4164 & 912 & The Latin American Brain Health Institute dataset comprises multimodal data of 780 participants from Latin American with neurodegenerative diseases.\\
    \midrule
    \makecell{Chisco~\citep{zhang2024chisco}} & \makecell{Imagined\\Speech}  & 62 & 30 & 9941 & 2061 & Chinese Imagined Speech Corpus (Chisco), including over 20,000 sentences of high-density EEG recordings of imagined speech from healthy adults.\\
    \midrule
    \makecell{Things EEG 1\\\citep{grootswagers2022things-eeg-1}} & \makecell{Visual\\Stimulus}  & 62 & 20 & 6568 & 1749 & The THINGS-EEG dataset provides neuroimaging recordings to a systematic collection of objects and concepts. \\
    \midrule
    \makecell{OpenMIIR~\citep{stober2015open-miir}} & \makecell{Acoustic\\Stimulus}  & 64 & 10 & 3399 & 992 & A public domain dataset of EEG recordings taken during music perception and imagination comprised 10 subjects listening to and imagining 12 short music fragments. \\
    \bottomrule
    \end{tabular}
    }

\end{table}

\paragraph{SEED-family datasets.}
The pretraining corpus includes SEED-IV, SEED-GER, and SEED-FRA, while SEED and SEED-V are used as downstream emotion-recognition datasets~\citep{liu2022seed-v}. SEED and SEED-IV were collected from different participants~\citep{yi2023mmm}. SEED-GER and SEED-FRA extend the SEED family to German and French populations, with eight participants in each dataset~\citep{liu2022identifying}. The downstream SEED and SEED-V results follow the subject-dependent trial-splitting protocols described in Appx.~\ref{appx:exp_setting}.

\subsection{Fine-tuning Datasets}
The detailed information on the evaluation datasets are listed in Tab.~\ref{tab:finetune-dataset}: 
(1) Seizure Detection: \textbf{Siena}~\citep{detti2020siena} (binary classification with seizure or healthy); 
(2) Emotion Recognition: \textbf{SEED}~\citep{zheng2015seed} (3-class classification with sad, neutral or happy), \textbf{SEED-V}~\citep{liu2022seed-v} (5-class classification with disgust, fear, sad, neutral or happy); 
(3) Motor Imagery: \textbf{PhysioMI}~\citep{Schalk2004physiomi} (4-class classification with left fist, right fist, both fists or feet), \textbf{Mimul-11}~\citep{jeong2020mimul-11} (3-class classification with reaching, grasping or twisting); 
(4) Mental Stress: \textbf{Workload}~\citep{zyma2019workload} (binary-class classification with arithmetic calculation or resting); 
(5) Sleep Staging: \textbf{HMC}~\citep{alvarez2021hmc} (5-class classification with wake, REM, N1, N2 or N3); 
(6) Anomalous Event Detection: \textbf{TUEV}~\citep{harati2015tuev} (6-class classification with spike and slow wave, generalized periodic epileptiform discharge, periodic lateralized epileptiform dischage, eye movement artifact or background); 
(7) Abnormal Classification: \textbf{TUAB}~\citep{lopez2015tuab} (binary-class classification with abnormal or normal); 
(8) Visual Target Detection: \textbf{Things-EEG-2}~\citep{gifford2022things-eeg-2} (binary-class classification with target or non-target); 
(9) Alzheimer 's Disease Identification: \textbf{ADFTD}~\citep{miltiadous2023adftd} (3-class classification with Alzheimer's Disease, Frontotemporal Dementia or healthy); 
(10) Slowing Event Classification: \textbf{TUSL}~\citep{weltin2017tusl} (3-class classification with seizure, slow wave or background).

\begin{table}[ht]
    \centering
    \caption{Detailed information about evaluation datasets.}
    \label{tab:finetune-dataset}
    \resizebox{1.0\textwidth}{!}{
    \begin{tabular}{cccccccc}
    \toprule
    Dataset     &  Category    & \#Channel & Duration & \#Train & \#Valid & \#Test & Task \\
    \midrule
    TUAB            & Abnormal Classification       & 23    & 30    & 247728    & 12315     & 12277     &  Binary Classification  \\
    TUEV            & Anomalous Event Detection     & 21    & 5     & 87834     & 12473     & 13046     &  6-class Classification \\
    TUSL            & Slowing Event Classification  & 21,22 & 10    & 210       & 43        & 37        &  3-class Classification \\
    SEED            & Emotion Recognition           & 60    & 10    & 22455     & 7875      & 7560      &  3-class Classification \\
    SEED-V          & Emotion Recognition           & 60    & 10    & 3552      & 4638      & 4128      &  5-class Classification \\
    HMC             & Sleep Staging                 & 4     & 30    & 91681     & 22804     & 22440     &  5-class Classification \\
    Workload        & Mental Stress                 & 19    & 10    & 1537      & 300       & 297       &  Binary Classification  \\
    Siena           & Seizure Detection             & 29    & 10    & 41631     & 5592      & 3607      &  Binary Classification  \\
    Mimul-11        & Motor Imagery                 & 60    & 5     & 31398     & 5000      & 4949      &  3-class Classification \\
    PhysioMI        & Motor Imagery                 & 64    & 4     & 6210      & 1734      & 1803      &  4-class Classification \\
    Things-EEG-2    & Visual Target Detection       & 63    & 5     & 24915     & 8324      & 8331      &  Binary Classification \\
    ADFTD           & \makecell{Alzheimer's Disease\\Identification}    & 19     & 10        & 4743      & 1115      & 1155      &  3-class Classification \\
    \bottomrule
    \end{tabular}
    }

\end{table}

\section{Implementation Details}\label{appx:implementation}

\subsection{Overview}
Let the raw EEG signal be denoted as $\mathbf{x}_0 \in \mathbb{R}^{C \times P_{0}}$, where $C$ is the number of channels and $P_{0}$ is the total number of time points.
This signal is segmented into non-overlapping patches of length $P = f_s$.
After batching, the batched input tensor is defined as $\mathbf{x}_p \in \mathbb{R}^{B \times T \times C \times P}$, where $B$ is the batch size and $T$ is the number of temporal patches.
Throughout this paper, $D_{m}$ denotes the model embedding dimension, $\bigoplus$ represents the feature concatenation operator, and $\text{RMS}$ and $\text{ZN}$ refer to Root Mean Square and Z-score normalization, respectively.

Our framework consists of a shared encoder and two distinct decoder configurations.
The encoder includes: (1) a \textbf{Prior-Guided Continuous Tokenizer} and (2) a \textbf{Depth-Stratified Expert Transformer}.
The stage-specific components include (3) an \textbf{EEG Reconstructor} for pre-training and (4) \textbf{Dual-Stream Spatial-Temporal Classifiers} for fine-tuning.

\subsection{Prior-Guided Continuous Tokenizer}

\subsubsection{Feature Extraction and Positional Encoding}\label{appx:feature_extraction}
To capture temporal dynamics and global spectral characteristics, we extract features from both time and frequency domains.
In the temporal domain, we apply stacked 1D convolutional blocks with multiple receptive fields to each patch.
The normalized patch tensor $\mathbf{x}_p \in \mathbb{R}^{B \times T \times C \times P}$ is reshaped into $\mathbf{x}^\prime_p \in \mathbb{R}^{(BTC)\times 1 \times P}$ and processed via $S$ convolutional branches.
The aggregated temporal feature is defined as:
\begin{equation}
    \mathbf{x}_t = \mathrm{RMS}\left( \mathrm{GELU}\!\left(\mathbf{W}_{\mathrm{emb}}^\top \bigoplus_{i=1}^{S}\mathbf{Y}_i\right)\right),
    \label{eq:multi_scale_embed}
\end{equation}
where $\mathbf{Y}_i$ denotes the output of the $i$-th branch.
In the frequency domain, we extract spectral features using the Power Spectral Density~(PSD)~\citep{alsolamy2016emotion}.
Specifically, we apply Fast Fourier Transform~(FFT) to $\mathbf{x}_p$ to obtain its frequency representation.
The one-sided PSD is computed using a Hanning window and converted to the decibel scale, yielding the spectral feature $\mathbf{x}_s$.
The composite feature $\mathbf{x} \in \mathbb{R}^{B \times T \times C \times D_m}$ is formed by concatenating the normalized temporal and spectral features:
\begin{equation}
    \mathbf{x} = \bigoplus(\mathbf{ZN}(\mathbf{x}_t), \mathbf{ZN}(\mathbf{x}_s)).
    \label{eq:embed_concat}
\end{equation}
For the pre-training reconstruction task, the target signal $\mathbf{x}_{\text{ref}}$ consists of the concatenated raw patch $\mathbf{x}_p$ and spectral features $\mathbf{x}_{s}$.

EEG channel configurations vary significantly across datasets.
To address this, we define a standardized electrode superset $\mathbb{C}$ (based on the 10-10 system with auxiliary electrodes such as T1 and T2) and introduce a learnable channel positional embedding $\mathbf{E}_c\in\mathbb{R}^{|\mathbb{C}|\times D_m}$, which is added to the input features $\mathbf{x}$ and serves as a query in the EEG reconstructor.
For the temporal dimension, we apply Rotary Positional Encoding~(RoPE) within the attention to encode relative temporal positions.

\subsubsection{Static Group Queries with Weak Sensor-Space Priors}\label{appx:static_prior}
A core strategy of our design involves representing EEG signals not only as channel patches but also as a set of \emph{group patches} corresponding to neuro-functional organization.
We define a fixed grouping scheme $\mathcal{P}$ (details in Appx.~\ref{appx:elec_group}) that organizes channels according to anatomical regions and coarse functional networks.
To incorporate prior knowledge, we employ a prior matrix $\mathbf{M}_{\text{s}} \in \mathbb{R}^{|\mathcal{P}| \times C}$ that defines channel-group associations, wherein in-group entries are set to $0$ and out-of-group entries are assigned a negative bias~(ablations in Appx.~\ref{appx:extra_ablation}).
To prevent excessive rigidity, we introduce a learnable relaxation term $\mathbf{\Delta}_{\text{pri}} \in \mathbb{R}^{|\mathcal{P}| \times C}$, which adjusts this bias while retaining the underlying structural guidance.
Subsequently, we instantiate learnable parameters denoted as \emph{static queries} $Q_s\in\mathbb{R}^{|\mathcal{P}|\times D_m}$, which serve as inputs to the group cross-attention modules.
The attention mechanism incorporating these priors is formulated as:
\begin{equation}
    \mathrm{Attn}(Q_s, K, V) = \mathrm{Softmax}\left(\frac{Q_sK^\top}{\sqrt{d_k}} + \mathbf{M}_{\text{s}} + \mathbf{\Delta}_{\text{pri}}\right) V
    \label{eq:brain_attention_static_query}
\end{equation}
where $K$ and $V$ are from the input $\mathbf{x}$, reshaped to $\mathbb{R}^{(B \times T) \times C \times D_m}$.

\subsubsection{Dynamic Sub-second Queries through Short-Time Correlation}\label{appx:dynamic_prior}
Static groups provide coarse region-network anchors. Dynamic groups complement them with learnable short-time channel affinities that vary with the sample and temporal position within each one-second patch.
We reshape and embed each patch into \textbf{micro-frames}:
$\mathbf{x} \xrightarrow{} \tilde{\textbf{x}}\in\mathbb{R}^{B\times T\times F\times C\times S}, P=F\cdot S,$ 
\begin{align}
        E &=\phi(\tilde{\textbf{x}}) \in\mathbb{R}^{B\times T\times F\times C\times H},\\
        Q^\prime = E &W_Q, K^\prime = E W_K \space\in\mathbb{R}^{B\times T\times F\times C\times d},
\end{align}
where $F$, $S$, and $H$ are the number of frames, samples per frame, and frame embedding dimension, respectively; $\phi(\cdot)$ is a compact convolution projector, $W_Q, W_K \in \mathbb{R}^{H \times d}$ are projection matrices, and $d$ is the correlation feature dimension. 
For a given patch, the \textbf{channel affinity} between micro-frames $i$ and $j$ is defined as:
\begin{equation}
    A_{i,j} \;=\; \frac{1}{\sqrt d}\, Q^\prime_i K_j^{\prime\top}\;\;\in\mathbb{R}^{C\times C},
\end{equation}
where $Q^\prime_i, K^\prime_j \in \mathbb{R}^{C \times d}$ denote the slices of $Q^\prime$ and $K^\prime$ at frames $i$ and $j$, respectively; $A_{i,j}(a,b)$ quantifies the affinity between channel $a$ (at frame $i$) and channel $b$ (at frame $j$).
We define a lag-limited set of frame pairs,
    $\mathcal{G}=\{(i,j)\mid 0\le i-j\le \mathbf{\Delta}_f \},$
where $\mathbf{\Delta}_f$ denotes the maximum frame lag, and each pair $(i,j) \in \mathcal{G}$ constitutes a single dynamic group $g$.
The group biases are derived from each affinity matrix via a node encoder applied row-wise, $b_g \;=\; enc(A_g)\;\in\mathbb{R}^{C},$
where $A_g \in \mathbb{R}^{C \times C}$ denotes the affinity matrix for group $g$, and $enc(\cdot)$ is an MLP-based node encoder utilizing row statistics and top-$k$ neighbor values.
Aggregating these groups yields $\tilde{\textit{M}_{d}} \in \mathbb{R}^{B \times T \times |\mathcal{G}| \times C}$, which serves as an additive dynamic attention bias analogous to Eq.~\ref{eq:brain_attention_static_query}, but utilizes a distinct set of parameters denoted as \emph{dynamic queries} $\tilde{Q_d} \in \mathbb{R}^{|\mathcal{G}| \times D_m}$. To suppress uninformative lag pairs and reduce computation, we apply a \textbf{Gumbel Selector} to pick the most important frame-pairs in $\mathcal{G}$ for each sample-patch. We obtain differentiable importance scores with Gumbel-Softmax and select top-$k$ groups:
\begin{align}
    \mathbf{\Lambda} &= \mathrm{Mean_C}(\tilde{\textit{M}_{d}})+\mathrm{MLP}(\tilde{Q_d}) \in \mathbb{R}^{(B \cdot T) \times |\mathcal{G}|} \\
    \Pi &= \operatorname{TopK}(\mathrm{GumbelSoftmax}(\mathbf{\Lambda};\tau), k) \in \mathbb{R}^{(B \cdot T) \times k}
\end{align}
where $\tau$ is temperature. The $\Pi$ score is used to gather corresponding dynamic biases $\mathbf{M}_d \in \mathbb{R}^{B \times T \times k\times C} $ and queries $ Q_d \in \mathbb{R}^{k\times D_m} $.

\subsection{CKA-Calibrated Depth-Stratified Expert Transformer}
\label{appx:cka_moe}
During pretraining, PRiSE-EEG contains an ordered MoE Transformer stack consisting of tokenizer cross-attention blocks, temporal-wise blocks, group-wise blocks, and decoder blocks.
For PRiSE-EEG-B, this stack has $L=12$ blocks: 2 tokenizer cross-attention blocks, 4 group MoE blocks, 4 temporal MoE blocks, and 2 decoder MoE blocks.
All blocks use the same MoE-FFN structure, so CKA-calibrated shared expert allocation can be applied uniformly across the stack.

\paragraph{Interleaved temporal and group layers.}
Given the tokenizer output $\mathbf{h}_{0}\in\mathbb{R}^{B\times T\times G\times D_m}$, the encoder contains $L_E$ Transformer layers. Temporal-wise and group-wise Transformer layers appear alternately, so each type has $L_E/2$ layers~\citep{bertasius2021timesformer,arnab2021vivit,li2025estformer}. For layer $l$, let $a_l\in\{\mathrm{temp},\mathrm{group}\}$ denote its attention type. If $a_l=\mathrm{temp}$, self-attention is applied along the temporal dimension $T$ for each group. If $a_l=\mathrm{group}$, self-attention is applied along the group dimension $G$ for each time patch. Each temporal-wise or group-wise layer has its own depth-stratified MoE feed-forward block:
\begin{equation}
    \tilde{\mathbf{h}}_l
    =
    \mathrm{Attn}^{a_l}_l(\mathbf{h}_l),
    \qquad
    \mathbf{h}_{l+1}
    =
    \tilde{\mathbf{h}}_l
    +
    \mathrm{DSE}_l(\mathrm{LN}(\tilde{\mathbf{h}}_l)).
\end{equation}
Thus, depth stratification is applied to the FFN of every Transformer layer, regardless of whether the layer performs temporal-wise or group-wise attention.

\paragraph{Dense pilot checkpoints.}
Before training PRiSE-EEG, we train a dense pilot version with the same tokenizer and interleaved temporal/group attention layout, but with standard dense FFNs. We save pilot checkpoints at selected pretraining steps $\mathcal{T}=\{t_1,\dots,t_m\}$. At each step $t\in\mathcal{T}$, we feed a mixed-paradigm batch $\mathcal{B}_t=\{(\mathbf{x}_i,y_i)\}_{i=1}^{N_b}$, where $y_i$ denotes the dataset or paradigm identity of sample $i$.

\paragraph{Sample-wise CKA.}
For block $l$, let $N_l$ denote the number of tokens in that block representation.
For sample $i$ at pretraining step $t$, we reshape the block output as
\begin{equation}
    \mathbf{R}_{l,t}^{(i)}=\mathrm{reshape}(\mathbf{h}_{l,t}^{(i)})\in\mathbb{R}^{N_l\times D_m}
\end{equation}
be its flattened token representation. We use linear CKA between two samples $i$ and $j$:
\begin{equation}
    \mathrm{CKA}_{l,t}(i,j)
    =
    \frac{
    \left\|
    (\bar{\mathbf{R}}_{l,t}^{(i)})^\top
    \bar{\mathbf{R}}_{l,t}^{(j)}
    \right\|_F^2
    }{
    \left\|
    (\bar{\mathbf{R}}_{l,t}^{(i)})^\top
    \bar{\mathbf{R}}_{l,t}^{(i)}
    \right\|_F
    \left\|
    (\bar{\mathbf{R}}_{l,t}^{(j)})^\top
    \bar{\mathbf{R}}_{l,t}^{(j)}
    \right\|_F
    +
    \epsilon
    },
\end{equation}
where $\bar{\mathbf{R}}=\mathbf{J}\mathbf{R}$, $\mathbf{J}_{l}=\mathbf{I}_{N_l}-\frac{1}{N_l}\mathbf{1}\mathbf{1}^{\top}$ centers the token dimension, $\|\cdot\|_F$ is the Frobenius norm, and $\epsilon$ is a small constant.

\paragraph{Within- and between-paradigm CKA.}
Let $\mathcal{Y}$ be the set of paradigms in the mixed batch. For paradigm $a$, define within-paradigm pairs
\begin{equation}
    \mathcal{I}_{a,a}
    =
    \{(i,j)\mid i<j,\; y_i=a,\; y_j=a\}.
\end{equation}
For two different paradigms $a$ and $b$, define between-paradigm pairs
\begin{equation}
    \mathcal{I}_{a,b}
    =
    \{(i,j)\mid y_i=a,\; y_j=b\}.
\end{equation}
We first average over pairs inside each paradigm or paradigm pair, and then average over paradigms:
\begin{equation}
    c^{\mathrm{intra}}_{l,t}
    =
    \frac{1}{|\mathcal{Y}|}
    \sum_{a\in\mathcal{Y}}
    \frac{1}{|\mathcal{I}_{a,a}|}
    \sum_{(i,j)\in\mathcal{I}_{a,a}}
    \mathrm{CKA}_{l,t}(i,j),
\end{equation}
\begin{equation}
    c^{\mathrm{inter}}_{l,t}
    =
    \frac{2}{|\mathcal{Y}|(|\mathcal{Y}|-1)}
    \sum_{\substack{a,b\in\mathcal{Y}\\a<b}}
    \frac{1}{|\mathcal{I}_{a,b}|}
    \sum_{(i,j)\in\mathcal{I}_{a,b}}
    \mathrm{CKA}_{l,t}(i,j).
\end{equation}
Here $c^{\mathrm{intra}}_{l,t}$ measures within-paradigm representation consistency at layer $l$ and pretraining step $t$, while $c^{\mathrm{inter}}_{l,t}$ measures cross-paradigm alignment.

\paragraph{CKA-derived shared expert ratio.}
We define normalized sharedness as
\begin{equation}
    s_{l,t}
    =
    \mathrm{clip}
    \left(
    \frac{c^{\mathrm{inter}}_{l,t}}
    {c^{\mathrm{intra}}_{l,t}+\epsilon},
    0,1
    \right),
    \qquad
    \bar{s}_l
    =
    \frac{1}{|\mathcal{T}|}
    \sum_{t\in\mathcal{T}}s_{l,t}.
\end{equation}
The shared expert proportion of layer $l$ is
\begin{equation}
    \rho_l^{\mathrm{sh}}
    =
    \rho_{\min}
    +
    (\rho_{\max}-\rho_{\min})
    \sigma
    \left(
    \frac{\bar{s}_l-\tau}{T_{\mathrm{cka}}}
    \right),
\end{equation}
where $\rho_{\min}$ and $\rho_{\max}$ bound the shared-expert proportion, $\tau$ is the sharedness threshold, and $T_{\mathrm{cka}}$ is the sigmoid temperature. Here $E_l$ is the total number of experts, including one fixed shared expert, and $\rho_l^{\mathrm{sh}}$ is the total shared proportion. We set
\begin{equation}
    E_l^{\mathrm{sh}}
    =
    \mathrm{round}(E_l\rho_l^{\mathrm{sh}}),
    \qquad
    E_l^{\mathrm{sp}}
    =
    E_l-E_l^{\mathrm{sh}}.
\end{equation}
The allocation is computed once from the dense pilot pretraining steps and then fixed during PRiSE-EEG training.

\paragraph{Expert configuration and routing.}
Each layer of PRiSE-EEG-B contains one default shared expert and an eight-expert pool partitioned by CKA. The total shared count $E_l^{\mathrm{sh}}$ includes the default expert, leaving $E_l^{\mathrm{sh}}-1$ shared and $9-E_l^{\mathrm{sh}}$ specialized experts in the pool. With $\rho_{\min}=1/9$, $\rho_{\max}=5/9$, $\tau=0.275$, and $T_{\mathrm{cka}}=0.040$, the 12-layer allocation is:
\begin{equation}
[E_l^{\mathrm{sh}}]_{l=0}^{11}=[5,5,5,4,2,1,1,2,2,2,2,2].
\end{equation}
The allocation is determined before PRiSE-EEG training and remains fixed thereafter. For token $\mathbf{h}_{l,n}$, the router assigns logits and probabilities only to the specialized expert pool:
\begin{align}
a_{l,e}(\mathbf{h}_{l,n})
&=\mathbf{w}_{l,e}^{\top}\mathbf{h}_{l,n},\\
p_{l,e}(\mathbf{h}_{l,n})
&=\frac{\exp(a_{l,e}(\mathbf{h}_{l,n}))}
{\sum_{j\in\mathcal{E}_l^{\mathrm{sp}}}\exp(a_{l,j}(\mathbf{h}_{l,n}))}.
\end{align}
The top-$K$ specialized experts, with $K=2$, are selected according to
\begin{equation}
\mathcal{K}_l(\mathbf{h}_{l,n})
=\operatorname{TopK}_{e\in\mathcal{E}_l^{\mathrm{sp}}}
\left(a_{l,e}(\mathbf{h}_{l,n})+b_{l,e},K\right).
\end{equation}
The output follows Eq.~\ref{eq:dse-output}: all shared experts process every token, whereas only the selected specialized experts are activated. Specialized-expert outputs are weighted by $p_{l,e}$ without renormalization over $\mathcal{K}_l$, and no token is dropped.

\paragraph{Load balancing.}
Following the DeepSeek-style bias-based balancing strategy~\citep{liu2024deepseek}, $b_{l,e}$ affects expert selection but not the routing probability. Let $N_l$ denote the number of tokens routed by layer $l$ in an update. The normalized load of specialized expert $e$ is
\begin{equation}
u_{l,e}=\frac{1}{K N_l}\sum_{n=1}^{N_l}
\mathbb{I}\!\left[e\in\mathcal{K}_l(\mathbf{h}_{l,n})\right].
\end{equation}
The selection bias is updated outside backpropagation as
\begin{equation}
b_{l,e}\leftarrow b_{l,e}-\eta_b
\left(u_{l,e}-\frac{1}{E_l^{\mathrm{sp}}}\right),
\label{eq:expert-bias-update}
\end{equation}
where $\eta_b = 10^{-4}$ is the expert-bias update rate.

We additionally use sequence-wise auxiliary balancing. For routing sequence $q$ with $N_{l,q}$ tokens:
\begin{align}
f_{l,q,e}
&=\frac{1}{K N_{l,q}}\sum_{n=1}^{N_{l,q}}
\mathbb{I}\!\left[e\in\mathcal{K}_l(\mathbf{h}_{l,q,n})\right],\\
\bar p_{l,q,e}
&=\frac{1}{N_{l,q}}\sum_{n=1}^{N_{l,q}}p_{l,e}(\mathbf{h}_{l,q,n}).
\end{align}
The layer-wise auxiliary loss is
\begin{equation}
\mathcal{L}_{\mathrm{aux},l}=\frac{E_l^{\mathrm{sp}}}{Q_l}
\sum_{q=1}^{Q_l}\sum_{e\in\mathcal{E}_l^{\mathrm{sp}}}
f_{l,q,e}\bar p_{l,q,e},
\label{eq:aux-layer}
\end{equation}
where $Q_l$ is the number of routing sequences in layer $l$.

\subsection{Decoder}

\subsubsection{EEG Reconstructor at Pre-training Stage}
During pre-training, the reconstructor maps the encoder context $\mathbf{h}$ to the channel space via a symmetric decoding mechanism employing learnable queries and Transformers.
The module reconstructs both waveform samples and spectral targets in $\mathbf{x}_{\text{ref}}$, with masked patches reconstruction and future patches forecasting objectives.

\subsubsection{Dual-Stream Average Pooling MLP Head at Fine-tuning Stage}
For unified fine-tuning, we append a compact classification head to the encoder context $\mathbf{h}\in\mathbb{R}^{B\times T\times G\times D_m}$.
We implement a dual-stream architecture that processes representations from two complementary perspectives:
(1) a temporal branch that averages over query groups to yield a tensor of shape $\mathbb{R}^{B\times T\times D_m}$, which is subsequently flattened and mapped to $\mathbb{R}^{B\times D_m}$ via an MLP; and
(2) a query branch that performs temporal averaging to produce a representation in $\mathbb{R}^{B\times G\times D_m}$, followed by a corresponding MLP.
The resulting features are concatenated, normalized, fused via a linear layer, and projected to task logits.
This configuration is designed to surpass the performance of standard average pooling while ensuring greater memory efficiency than fully flattened architectures.

\subsection{Pre-training and Fine-tuning Strategy}
\subsubsection{Pre-training Stage}\label{appx:pt_stage}
\noindent \textbf{Activation within Tokenizer.} During pre-training, to ensure comprehensive optimization of all group priors, we fully activate both static and dynamic group queries.

\noindent \textbf{Loss Function.} We pre-train PRiSE-EEG using a self-supervised framework comprising multiple objectives:
\begin{enumerate}[topsep=3pt,leftmargin=13pt]
    \item \textbf{Masked Auto-Encoding (MAE):} We randomly mask $50\%$ of the input along either temporal patches (MAE-TP) or channels (MAE-CH) and reconstruct the masked targets.
    \item \textbf{Generative Pre-training (GPT):} We train an auto-regressive objective to predict the next patch conditioned on previous patches via causal attention.
    \item \textbf{Auxiliary Sequence Routing Loss~(Aux):} To prevent expert collapse, we impose a load balancing loss that encourages uniform patch distribution across experts.
\end{enumerate}
We employ the reconstruction losses formulated as:
\begin{equation}
    \mathcal{L}_{\text{MAE}} = \frac{1}{|\Omega|} \sum_{i \in \Omega} \left\lVert\hat{\mathbf{x}}_{i} - \mathbf{x}_{\text{ref}, i}\right\rVert_2,
    \label{eq:mae-loss}
\end{equation}
where $\Omega$ denotes the set of masked patch indices, and
\begin{equation}
    \mathcal{L}_{\text{GPT}} = \frac{1}{C(T-1)} \sum_{n=1}^{C}\sum_{t=1}^{T-1} \left\lVert\hat{\mathbf{x}}_{n,t+1} - \mathbf{x}_{\text{ref}, n, t+1}\right\rVert_2.
    \label{eq:gpt-loss}
\end{equation}

The bias-based load correction is defined in Eq.~\ref{eq:expert-bias-update}. The sequence-wise terms in Eq.~\ref{eq:aux-layer} are averaged over the $L_E$ encoder layers:
\begin{equation}
    \mathcal{L}_{\mathrm{Aux}}=\frac{1}{L_E}
    \sum_{l=1}^{L_E}\mathcal{L}_{\mathrm{aux},l}.
    \label{eq:aux-loss}
\end{equation}
The overall pre-training loss is:
\begin{equation}
    \mathcal{L}_{pt} = \frac{1}{3} \left( \mathcal{L}_{\text{MAE-TP}} + \mathcal{L}_{\text{MAE-CH}} + \mathcal{L}_{\text{GPT}} \right) + \beta\mathcal{L}_{\text{Aux}}.
    \label{eq:pretrain-loss-appx}
\end{equation}
where $\beta$ governs the weight of the expert load balancing constraint.

\subsubsection{Unified Fine-tuning Stage}\label{appx:ft_stage}
\noindent \textbf{Activation within Tokenizer.} Recognizing that the relevance of specific group priors varies across tasks, we apply differentiable top-$k$ selection over the candidate dynamic groups using Gumbel-Softmax~\citep{jang2017categorical}, activating a subset of $k$ groups with their dynamic priors $\mathbf{M}_{d}$. 
The Gumbel temperature is annealed via an exponential decay to gradually sharpen the selection distribution.

\noindent \textbf{Loss Function.}
We fine-tune the shared pre-trained encoder jointly across multiple downstream datasets, equipping each task with a dedicated \textbf{dual-stream classification head}.
For a set of $K$ tasks, the joint fine-tuning objective is defined as:
\begin{equation}
    \mathcal{L}_{ft} = \sum_{k=1}^{K} \mathcal{L}_{\text{cls}}^{k} + \beta\mathcal{L}_{\text{Aux}}.
\end{equation}
Each task employs a weighted cross-entropy loss:
\begin{equation}
    \mathcal{L}_{\text{cls}}^{k} = -\frac{1}{N} \sum_{i=1}^{N}\sum_{c=1}^{C_k} w_c^{k} y_{ic}^{k} \log(p_{\theta}(y_{ic}^{k} | \mathbf{x}_i)),
\end{equation}
where $N$ is the batch size, $C_k$ is the number of classes for task $k$, $w_c^{k}$ is the class weight, $y_{ic}^{k}$ is the ground truth label, and $p_{\theta}$ is the predicted probability.
This unified strategy facilitates cross-dataset transfer and enhances data efficiency within a shared encoder.

\section{Gradient and Routing Analysis}
\label{appx:gradient_analysis}

We analyze encoder gradients to quantify whether different downstream datasets impose aligned or conflicting updates on the same shared encoder. Let $\mathcal{D}=\{D_1,\dots,D_M\}$ be the downstream datasets. Starting from a fixed pretrained checkpoint, we attach dataset-specific classification heads and run $Q$ consecutive optimization steps for each dataset without updating the checkpoint used for other datasets. Unless otherwise stated, $Q=512$.

\subsection{Projected Encoder Gradients}
We partition encoder parameters into semantic groups $g\in\mathcal{G}$, such as tokenizer, temporal layers, group layers, or backbone FFNs. At step $q$ for dataset $D_i$, let $\mathbf{g}_{g,i}^{(q)}$ be the flattened encoder gradient of group $g$. To reduce memory cost, we project gradients into $\mathbb{R}^{d}$ with a fixed CountSketch~\citep{CHARIKAR20043countsketch} projector $\mathbf{S}_g$:
\begin{equation}
    \mathbf{v}_{g,i}^{(q)}
    =
    \mathbf{S}_g
    \mathbf{g}_{g,i}^{(q)}
    \in
    \mathbb{R}^{d}.
\end{equation}
The same projector $\mathbf{S}_g$ is used for all datasets and steps within group $g$.

\subsection{Mean Gradient Cosine Similarity}
For each group $g$ and dataset $D_i$, we compute the normalized mean update direction
\begin{equation}
    \bar{\mathbf{v}}_{g,i}
    =
    \frac{
    \sum_{q=1}^{Q}\mathbf{v}_{g,i}^{(q)}
    }{
    \left\|
    \sum_{q=1}^{Q}\mathbf{v}_{g,i}^{(q)}
    \right\|_2+\epsilon
    }.
\end{equation}
The pairwise mean gradient cosine similarity is
\begin{equation}
    C_g(i,j)
    =
    \left\langle
    \bar{\mathbf{v}}_{g,i},
    \bar{\mathbf{v}}_{g,j}
    \right\rangle.
\end{equation}
Positive values indicate aligned average update directions, while negative values indicate conflicting mean directions. When reporting an overall heatmap, we average over groups:
\begin{equation}
    C(i,j)=\frac{1}{|\mathcal{G}|}\sum_{g\in\mathcal{G}}C_g(i,j).
\end{equation}

\subsection{Gradient Subspace Affinity}
Mean directions can miss cases where two datasets share dominant update modes but have different average gradients. Therefore, we also compare low-rank gradient subspaces. For group $g$ and dataset $D_i$, we stack projected gradients into
\begin{equation}
    \mathbf{X}_{g,i}
    =
    \begin{bmatrix}
    (\mathbf{v}_{g,i}^{(1)})^\top\\
    \cdots\\
    (\mathbf{v}_{g,i}^{(Q)})^\top
    \end{bmatrix}
    \in
    \mathbb{R}^{Q\times d}.
\end{equation}
Let $\mathbf{U}_{g,i}\in\mathbb{R}^{d\times k}$ be the top-$k$ PCA basis of $\mathbf{X}_{g,i}$ with orthonormal columns. We use $k=7$ by default. For datasets $D_i$ and $D_j$, the subspace affinity is
\begin{equation}
    \mathrm{Affinity}_g(i,j)
    =
    \frac{1}{k}
    \sum_{\ell=1}^{k}
    \sigma_{\ell}
    \left(
    \mathbf{U}_{g,i}^{\top}\mathbf{U}_{g,j}
    \right),
\end{equation}
where $\sigma_{\ell}(\cdot)$ is the $\ell$-th singular value. Values close to $1$ indicate highly overlapping dominant gradient subspaces, while values close to $0$ indicate distinct update subspaces. The overall affinity is
\begin{equation}
    \mathrm{Affinity}(i,j)
    =
    \frac{1}{|\mathcal{G}|}
    \sum_{g\in\mathcal{G}}
    \mathrm{Affinity}_g(i,j).
\end{equation}

\subsection{Expert Routing Similarity}
To compare gradient geometry with expert specialization, we also compute the cumulative routing distribution of each dataset. Let $\mathbf{p}_i\in\mathbb{R}^{E}$ be the normalized count or probability mass assigned to all experts by samples from dataset $D_i$. The routing similarity is
\begin{equation}
    R(i,j)
    =
    \frac{
    \langle \mathbf{p}_i,\mathbf{p}_j\rangle
    }{
    \|\mathbf{p}_i\|_2\|\mathbf{p}_j\|_2+\epsilon
    }.
\end{equation}
Expert routing similarity are then averaged across encoder Transformer layers. Comparing $\mathrm{Affinity}(i,j)$ and $R(i,j)$ shows whether datasets with similar dominant gradient subspaces also induce similar expert usage.

\section{Experimental Setup}\label{appx:exp_setting}

\subsection{Data Preprocessing}
The preprocessing pipeline utilizes a systematic framework for managing heterogeneous EEG data using MNE-Python~\citep{Gramfort2013mne}.
The pipeline begins with a resampling stage to transform all source signals to $f_s=256$ Hz, which standardizes patch division and ensures parity with the sampling rates used in baseline models.
A band-pass Finite Impulse Response (FIR) filter (1-100 Hz), implemented via an overlap-add method, is applied to attenuate low-frequency noise and high-frequency artifacts. This method ensures filtering consistency across signals of varying durations.
A notch filter is then applied at either $50$ Hz or $60$ Hz to remove power-line interference, with the frequency selected according to a manual review of each dataset's spectral profile or geographical origin.
Then time series are segmented into consecutive 1-second patches.
Electrode configurations are then mapped to a standardized 10-10 universal channel set. Additionally, data units are converted from microvolts ($\mathrm{\mu V}$) to Volts to ensure compatibility with MNE-Python conventions.
To optimize data handling, processed EEG signals are serialized in the \texttt{Parquet} format using Zstandard compression for storage efficiency, or \texttt{Apache Arrow} for high-throughput memory mapping.
To facilitate large-scale distributed training, the pipeline supports direct dataset access from remote object storage via the S3 protocol.
The implementation employs parallel execution to process the multi-terabyte pre-training corpus efficiently.

\subsection{Comparing Baselines}
We evaluate our framework against seven state-of-the-art EEG foundation models: BENDR~\citep{kostas2021bendr}, BIOT~\citep{yang2023biot}, LaBraM~\citep{jiang2024large}, EEGPT~\citep{wang2024eegpt}, CBraMod~\citep{wang2024cbramod}, CSBrain~\citep{zhou2025csbrain}, and REVE~\citep{Ouahidi2025reve}.
BENDR~\citep{kostas2021bendr} ($3.97$\,M parameters) applies a BERT-style objective to a large-scale clinical EEG corpus. BIOT~\citep{yang2023biot} ($3.19$\,M) manages data heterogeneity by tokenizing various biosignals into a unified, sequence-based structure.
LaBraM ($5.82M$) is pre-trained on 2,500 hours of data with integrated VQ-VAE~\citep{van2017neural} modules for dual-domain (frequency/phase) mask learning, EEGPT ($25.29M$) combines dual self-supervised universal representation learning and stabilization mechanisms, and CBraMod~\citep{wang2024cbramod} ($4.92M$) utilizes criss-cross attention to capture the spatial-temporal separately at the same transformer layer, leading to refined representation ability.
CSBrain~\citep{zhou2025csbrain} ($8.86M$) employs cross-scale spatiotemporal tokenization (CST) and structured sparse attention to obtain robust EEG representations.
REVE~\citep{Ouahidi2025reve} ($69.19M$) is characterized by the use of the largest pre-training dataset available and 4D Fourier positional encoding, which is paired with an improved MAE reconstruction objective.
In addition, we select two strong supervised methods for comparison, and two supervised methods (EEGNet~\citep{lawhern2018eegnet}, EEGConformer~\citep{song2022eegconformer}).
Among these methods, BENDR, BIOT, LaBraM, CBraMod, CSBrain and REVE exploit fully flatten MLP and fine-tune on different downstream tasks separately during the evaluation. 
However, EEGPT exploits linear probe to frozen pre-trained backbone, making their framework support the multi-task evaluation with different linear probes.
During experiment, we follow their own strategy and open-sourced hyperparameter settings to reproduce them on 12 downstream tasks based on their pre-trained EEG foundation models.

\subsection{Evaluation Strategy}
To partition each dataset, we perform a \textbf{subject-level split} for most of the datasets. Subject-dependent strategy is adopted for SEED and SEED-V to have a comparable metric to other baselines.
We also employ a \textbf{greedy, multi-label stratified} splitting algorithm to ensure label distributions remain balanced across the training, validation, and test sets to predefined ratios:
\begin{itemize}[topsep=0pt,leftmargin=13pt]
    \item \textbf{Siena}: Subject 0-7, 9-13, 16-17 are assigned to the training, valid, test set correspondingly;
    \item \textbf{SEED}: Following prior research, the 15 trials are divided into three sets in a 9:3:3 ratio, and all sessions are merged together; 
    \item \textbf{SEED-V}: Following prior research, the 15 trials are divided into three sets in a 1:1:1 ratio;
    \item \textbf{PhysioMI}: Subject 1-70,71-89, 90-109 are assigned to the training set, valid set, test set correspondingly;
    \item \textbf{Mimul-11}: Stratified splitting is employed to achieve ratios of 0.76, 0.12, and 0.12 for training, validating, and testing;
    \item \textbf{Workload}: Stratified splitting is employed to achieve ratios of 0.72, 0.14, and 0.14 for training, validating, and testing;
    \item \textbf{HMC}: Subjects are randomly split into training, validation, and test sets at a ratio of 103:24:24;
    \item \textbf{TUEV} and \textbf{TUSL}: Owing to the highly imbalanced label distribution in these datasets, the stratified splitting function is employed to create three splits from all the data, approximately aligning with predefined ratios of 0.8, 0.1, and 0.1; 
    \item \textbf{TUAB}: The validation and test sets are obtained by equally splitting the original evaluation set by subject, while the training set remains unchanged; 
    \item \textbf{Things-EEG-2}: Split by subject with predefined ratios of 0.6, 0.2, and 0.2;
    \item \textbf{ADFTD}: Stratified splitting is employed to achieve ratios of 0.70, 0.15, and 0.15 for training, validating, and testing.
\end{itemize}

\begin{table}[ht]
    \centering
    \caption{Configurations for model variants.}
    \label{tab:conf-variant}
    \resizebox{1.0\textwidth}{!}{
    \begin{tabular}{@{}lccccc@{}}
    \toprule
    & PRiSE-EEG-S & PRiSE-EEG-M & PRiSE-EEG-B & PRiSE-EEG-L & PRiSE-EEG-XL \\
    \midrule
    \#Parameters                    & 9.41 M & 27.06 M & 41.06 M & 78.86 M & 125.58 M \\
    \#Activated Parameters          & 6.46 M & 11.73 M & 18.05 M & 36.39 M & 64.85 M \\
    Model Dim                       & 192 & 256 & 384 & 512 & 640 \\
    MoE Expert Dim                  & 256 & 256 & 256 & 256 & 256 \\
    \makecell[l]{Tokenizer Cross Attention Block}    & 1 & 2 & 2 & 2 & 2 \\
    \makecell[l]{Group MoE Block}       & 3 & 3 & 4 & 6 & 8 \\
    \makecell[l]{Temporal MoE Block}    & 3 & 3 & 4 & 6 & 8 \\
    \makecell[l]{Decoder Block}         & 1 & 2 & 2 & 3 & 4 \\
    Batch Size                      & 256 & 128 & 128 & 128 & 64 \\
    Attn. Head                      & 4 & 4 & 8 & 8 & 8 \\
    \bottomrule
    \end{tabular}
    }
\end{table}

\subsection{Evaluation Metrics}
To account for class imbalance in the downstream datasets, we utilize the following evaluation metrics:
\begin{itemize}[topsep=0pt,leftmargin=13pt]
    \item \textbf{Balanced Accuracy}: the arithmetic mean of recall (sensitivity) across all classes, mitigating the impact of imbalanced class distributions. It is effective for evaluating classification models on datasets with significant disparities in class proportions, which is formulated as:
    \begin{align}
        \text{B-Acc} = \frac{1}{C} \sum_{i=1}^{C} \frac{TP_i}{TP_i + FN_i},
    \end{align}
    where $C$ is the number of classes, $TP_i$ and $FN_i$ denote true positives and false negatives for class $i$.
    \item \textbf{Weighted F1}: a harmonic mean of precision and recall, weighted by the number of true instances in each class. This metric accounts for class imbalance by assigning higher importance to classes with larger sample sizes, ensuring a more representative evaluation of model effectiveness, which is formulated as
    \begin{align}
        \text{Pre}_i &= \frac{TP_i}{TP_i + FP_i}, \quad \text{Rec}_i = \frac{TP_i}{TP_i + FN_i}, \\
        \text{W-F1} &= \sum_{i=1}^{C} w_i \cdot \frac{2 \cdot \text{Pre}_i \cdot \text{Rec}_i}{\text{Pre}_i + \text{Rec}_i},
    \end{align}
    where $FP_i$ denotes false positives for class $i$, and $w_i$ is the weight of class $i$ based on its support.
    \item \textbf{AUROC}: area under the ROC curve. It reflects the model’s ability to discriminate between classes across all possible decision boundaries, which is formulated as
    \begin{align}
        \text{TPR} = \frac{TP}{TP + FN}, \quad \text{FPR} = \frac{FP}{FP + TN}, \\
        \text{AUROC} = \int_{0}^{1} \text{TPR}(f) \, df, \quad f = \text{FPR}.
    \end{align}
    \item \textbf{AUC-PR}: area under the precision-recall curve. It provides a holistic evaluation of model performance under class imbalance, which is formulated as
    \begin{align}
        \text{AUC-PR} = \int_{0}^{1} \text{Pre}(r) \, dr, \quad r = \text{Rec}.
    \end{align}
    \item \textbf{Cohen’s Kappa}: the agreement level between predicted and true labels by comparing observed and expected frequencies along the diagonal of a confusion matrix. It is particularly suited for multi-class classification scenarios, which is formulated as
    \begin{align}
        \kappa = \frac{p_o - p_e}{1 - p_e},
    \end{align}
    where $p_o$ and $p_e$ are observed and expected agreement.
\end{itemize}
Among these metrics, AUROC and AUC-PR are used to evaluate binary classification tasks, while Cohen's Kappa and Weighted F1 are applied to multi-category classification. 
Together, these metrics provide a robust evaluation framework under class imbalance.



\begin{table}[tp]
\centering
\caption{Region electrode groupings used in the Prior-Guided Continuous Tokenization.}
\label{tab:region_groups}
\begin{tabular}{@{}cp{0.8\linewidth}@{}}
\toprule
\textbf{Group} & \textbf{Electrodes} \\
\midrule
Prefrontal & FP1, FPZ, FP2, AF9, AF7, AF5, AF3, AF1, AFZ, AF2, AF4, AF6, AF8, AF10 \\ \addlinespace
Frontal & F9, F7, F5, F3, F1, FZ, F2, F4, F6, F8, F10, FT9, FT7, FC5, FC3, FC1, FCZ, FC2, FC4, FC6, FT8, FT10 \\ \addlinespace
Central & FC1, FCZ, FC2, C5, C3, C1, CZ, C2, C4, C6, CP1, CPZ, CP2 \\ \addlinespace
Parietal & TP9, TP7, CP5, CP3, CP1, CPZ, CP2, CP4, CP6, TP8, TP10, P9, P7, P5, P3, P1, PZ, P2, P4, P6, P8, P10 \\ \addlinespace
Occipital & PO9, PO7, PO5, PO3, PO1, POZ, PO2, PO4, PO6, PO8, PO10, O1, OZ, O2, I1, IZ, I2 \\ \addlinespace
Left Temporal & T1, T7, T9, FT7, FT9, TP7, TP9 \\ \addlinespace
Right Temporal & T2, T8, T10, FT8, FT10, TP8, TP10 \\ \addlinespace
Midline & FPZ, AFZ, FZ, FCZ, CZ, CPZ, PZ, POZ, OZ, IZ \\
\bottomrule
\end{tabular}

\end{table}

\begin{table}[tp]
\centering
\caption{Network electrode groupings used in the Prior-Guided Continuous Tokenization.}
\label{tab:network_groups}
\begin{tabular}{@{}cp{0.8\linewidth}@{}}
\toprule
\textbf{Group} & \textbf{Electrodes} \\
\midrule
DMN & FP1, FPZ, FP2, AFZ, FZ, FCZ, CZ, CPZ, PZ, POZ, OZ, P5, P7, P6, P8, TP7, TP8, AF1, AF2 \\ \addlinespace
ECN & AF1, AF2, AF3, AF4, AF5, AF6, AF7, AF8, AF9, AF10, F1, F3, F5, F2, F4, F6, FC1, FC3, FC2, FC4, P1, P3, P2, P4, CP1, CP3, CP2, CP4 \\ \addlinespace
SN & FZ, FCZ, CZ, F1, F2, FC1, FC2, C1, C2, FC3, FC4, C3, C4 \\ \addlinespace
DAN & F1, F3, F2, F4, P1, P3, P5, P2, P4, P6, PO3, PO4 \\ \addlinespace
VAN & F10, F8, FT10, FT8, T8, TP8, C6, CP6, P10, P8, P6, T10, TP10 \\ \addlinespace
Visual & P9, P10, PO9, PO7, PO5, PO3, PO1, POZ, PO2, PO4, PO6, PO8, PO10, O1, OZ, O2, I1, IZ, I2, T10 \\ \addlinespace
Somatomotor & FC5, FC3, FC1, FC2, FC4, FC6, C5, C3, C1, CZ, C2, C4, C6, CP5, CP3, CP1, CP2, CP4, CP6 \\ \addlinespace
Language & F9, F7, F5, FT9, FT7, T7, TP7, C5, CP5, P9, P7, P5, T9, TP9 \\
\bottomrule
\end{tabular}
\end{table}

\begin{figure*}[t]
  \centering
  \includegraphics[width=\textwidth]{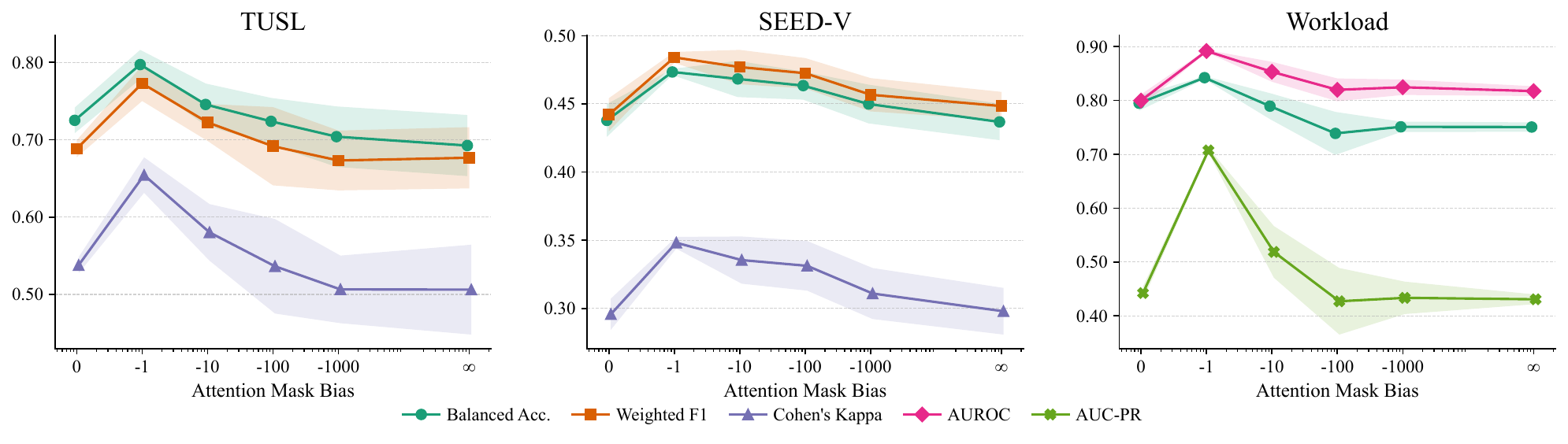}
  \caption{Ablation Study of attention bias strength for static path on TUSL, Workload and SEED-V datasets.}
  \label{fig:ablation-prior-mask}
\end{figure*}

\begin{figure*}[ht]
  \centering
  \includegraphics[width=\textwidth]{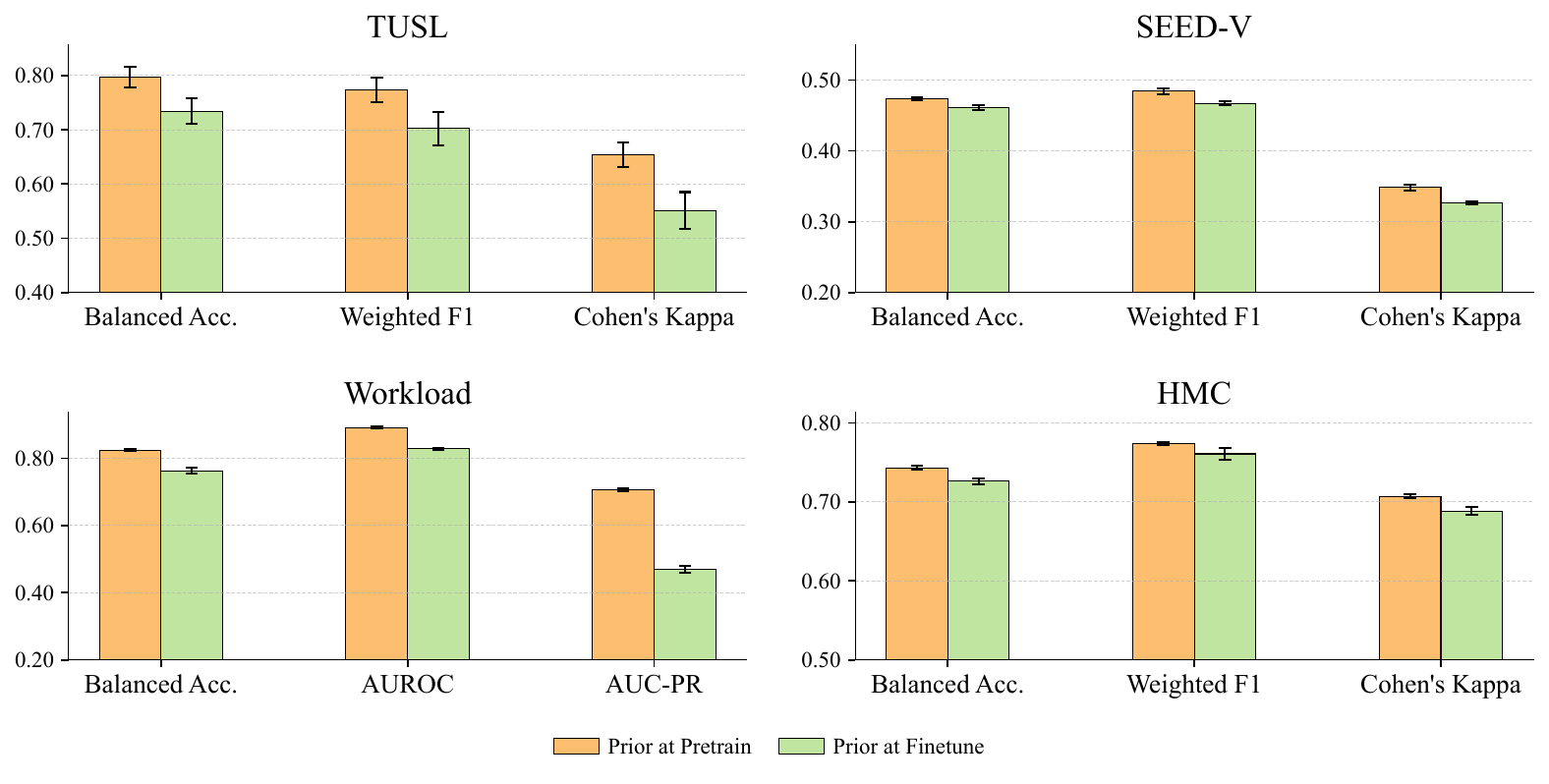}
  \caption{Ablation Study of the stage that the neural prior has been introduced on TUSL, SEED-V, HMC and Workload.}
  \label{fig:ablation-prior-stage}
\end{figure*}

\subsection{Training Settings}
\noindent\textbf{Data.} To facilitate high-throughput loading, all preprocessed samples are converted into the \texttt{Arrow} format, which accelerates distributed computation and enables direct data access for the GPU.

\noindent\textbf{Environment.} Experiments are conducted using Python 3.11.13, PyTorch 2.7.1, and CUDA 12.8 on four NVIDIA A100 GPUs for pre-training and one A100 GPU for fine-tuning. We utilize Automatic Mixed Precision (AMP) with the \texttt{bfloat16} data type to optimize memory usage and employ \texttt{GradScaler} to maintain stability.

\noindent\textbf{Optimizer.} We employ AdamW with a two-phase learning rate scheduler combining linear warm-up and cosine annealing. The warm-up factor is 0.1, and the gradient clipping threshold is 1.0. 
For MoE load balancing, the expert-bias update rate is $\eta_b=1\times10^{-4}$, and the auxiliary sequence balancing coefficient is $\beta=1\times10^{-5}$. For LoRA-based adaptation, the rank ($r$) and scaling factor ($\alpha$) are set to 8 and 16.0, respectively.

\noindent\textbf{Pre-training.} The pre-training learning rate is $1\times10^{-4}$, with attention and feedforward dropout rates set to 0.1. We apply a maximum \texttt{DropPath} rate of 0.1, utilizing a linear stochastic depth schedule that increases with layer depth. Training is conducted for 5 epochs, including a 1-epoch warm-up.

\noindent\textbf{Fine-tuning.} The fine-tuning learning rate is $5\times10^{-5}$ with a weight decay of 0.05. The process spans 20 epochs, consisting of one initial epoch with a frozen backbone to stabilize classifier initialization, followed by 2 warm-up epochs. \textbf{PRiSE-EEG} is optimized in a unified multi-task setting, where the shared backbone is updated using data from all downstream sets simultaneously. The classification head's learning rate is five times that of the backbone. The Gumbel selector is activated during this stage with $k=12$ and a temperature $\tau$ that decays from 1.0 to 0.1.
Model selection is based on peak validation performance, with final results reported on the held-out test set for each task.

\subsection{Model Configurations}\label{appx:model_config}
\textbf{PRiSE-EEG} is developed in five scaling variants, with parameter counts of 9.41\text{M}, 27.06\text{M}, 41.06\text{M}, 78.86\text{M}, and 125.58\text{M}.
The model scale is primarily adjusted by varying the number of Transformer blocks and the latent embedding dimensionality.
The patch size is set to $f_s=256$, corresponding to a 1-second temporal duration.
For the masked auto-encoding objectives, we employ a 50\% masking ratio applied independently to either temporal patches or entire channel sequences.
The hidden dimensionality of the dual-stream MLP classification head is 128. Detailed specifications for each model variant are provided in Tab.~\ref{tab:conf-variant}.

\subsection{Electrode Correspondence}\label{appx:elec_group}
The Prior-Guided Continuous Tokenization incorporates neuroscientific guidance by integrating priors from anatomical regions and functional brain networks.
Electrodes from the 10-10 standard montage are partitioned into groups as defined in Tabs.~\ref{tab:region_groups} and~\ref{tab:network_groups}.
Specifically, each electrode is assigned to its spatially proximal region of interest (ROI) within the reference atlas.
We utilize electrode coordinates registered in MRI space via the MNE-standard montage \citep{Gramfort2013mne}, with group correspondences visually verified against the Human Brainnetome Atlas \citep{fan2016human} using MRIcroGL \citep{rorden2025mricrogl}.
Within this partition scheme, certain electrodes are assigned to overlapping groups, reflecting their involvement in multiple functional networks or their location at anatomical boundaries.
While these groupings provide an initial neuro-informed prior, we acknowledge that these assignments are subject to future refinement as more granular electrophysiological mapping data becomes available.
The functional network groupings encompass the Default Mode Network~(DMN), Executive Control Network~(ECN), Salience Network~(SN), Dorsal Attention Network~(DAN), Ventral Attention Network~(VAN), Somatomotor Network, Visual Network, and Language Network.

\section{Extended Results}
\label{appx:extra_result}

\subsection{Detailed Experiment Results}
Tab.~\ref{tab:detailed-config-result} presents detailed model performance metrics across the default configuration and four alternative training strategies.
Multi-task unified fine-tuning achieves consistent improvements over single-task training, confirming that pre-trained representations provide a significant performance advantage compared to training from random initialization.
Besides, \textbf{PRiSE-EEG demonstrates strong representation transferability even with a frozen encoder}, validating the feature extraction capabilities of the Prior-Guided Continuous Tokenization and the CKA-Calibrated Depth-Stratified Expert Transformer.
LoRA-based fine-tuning performs worse than full-parameter fine-tuning under the tested configuration ($r=8$, $\alpha=16$). This suggests that the selected low-rank adaptation setting may provide insufficient flexibility for these tasks, rather than establishing a structural incompatibility between LoRA and MoE.

\begin{table}[htp]
\caption{Performance comparison of 5 configurations of PRiSE-EEG on 12 BCI tasks under \textbf{full-parameter unified}, \textbf{full-parameter single-task}, \textbf{efficient LoRA}, \textbf{freezing backbone} and \textbf{from scratch} fine-tuning.}
\label{tab:detailed-config-result}
\centering
\small
\resizebox{1.0\textwidth}{!}{
\setlength{\tabcolsep}{3.2mm}
\begin{tabular}{c c c c c c c c}
\toprule
\textbf{Dataset}   & \textbf{Metrics}  
              & \textbf{Unified} & \textbf{Single-Task} & \textbf{LoRA} & \textbf{Freezing} & \textbf{Scratch} & \textbf{Baseline Best} \\
\midrule

\multirow{3}{*}{\textbf{SEED}}
 & B-Acc       & \textbf{73.92$\pm$0.31} & 72.60$\pm$0.51  & 59.26$\pm$0.30  & 64.96$\pm$0.12  & 58.90$\pm$0.27  & 71.92$\pm$0.39 \\
 & F1/AUROC    & \textbf{73.29$\pm$0.22} & 72.00$\pm$0.62  & 58.90$\pm$0.28  & 64.66$\pm$1.04  & 58.68$\pm$0.27  & 71.84$\pm$0.20 \\
 & Kappa/AUCPR & \textbf{61.02$\pm$0.45} & 59.10$\pm$0.52  & 39.01$\pm$0.11  & 47.60$\pm$0.18  & 41.00$\pm$0.16  & 58.02$\pm$0.59 \\
\midrule

\multirow{3}{*}{\textbf{PhysioMI}}
 & B-Acc       & \textbf{65.55$\pm$0.38} & 63.53$\pm$1.01  & 32.62$\pm$0.14  & 41.81$\pm$1.20  & 30.08$\pm$1.26  & 64.80$\pm$0.53 \\
 & F1/AUROC    & \textbf{65.83$\pm$0.56} & 63.60$\pm$0.98  & 30.99$\pm$0.27  & 41.69$\pm$1.95  & 29.27$\pm$1.11  & 64.84$\pm$0.94 \\
 & Kappa/AUCPR & \textbf{54.78$\pm$0.69} & 52.10$\pm$1.02  & 10.17$\pm$0.17  & 22.40$\pm$0.91  & 6.79$\pm$0.82  & 53.04$\pm$0.85 \\
\midrule

\multirow{3}{*}{\textbf{Workload}}
 & B-Acc       & \textbf{82.47$\pm$0.30} & 77.33$\pm$1.53  & 59.70$\pm$0.40  & 74.96$\pm$1.85  & 74.74$\pm$1.17  & 74.23$\pm$1.36 \\
 & F1/AUROC    & \textbf{89.21$\pm$0.19} & 84.03$\pm$1.32  & 83.03$\pm$0.32  & 83.97$\pm$1.31  & 82.77$\pm$1.10  & 82.53$\pm$0.75 \\
 & Kappa/AUCPR & \textbf{70.75$\pm$0.47} & 68.17$\pm$1.47  & 62.66$\pm$0.43  & 62.16$\pm$1.30  & 61.09$\pm$0.95  & 61.48$\pm$1.90 \\
\midrule

\multirow{3}{*}{\textbf{TUEV}}
 & B-Acc       & \textbf{75.28$\pm$0.51} & 73.60$\pm$0.98  & 70.13$\pm$0.30  & 71.05$\pm$2.58  & 67.05$\pm$2.83  & 71.20$\pm$1.34 \\
 & F1/AUROC    & \textbf{88.25$\pm$0.51} & 87.90$\pm$0.79  & 81.87$\pm$0.57  & 79.87$\pm$3.18  & 79.03$\pm$1.57  & 85.07$\pm$0.85 \\
 & Kappa/AUCPR & \textbf{80.90$\pm$0.62} & 80.00$\pm$0.85  & 73.55$\pm$0.21  & 75.00$\pm$1.99  & 67.53$\pm$1.51  & 74.67$\pm$1.44 \\
\midrule

\multirow{3}{*}{\textbf{TUAB}}
 & B-Acc       & \textbf{82.22$\pm$0.12} & 81.83$\pm$0.52  & 81.11$\pm$0.31  & 81.39$\pm$0.29  & 78.38$\pm$1.10  & 81.75$\pm$0.65 \\
 & F1/AUROC    & \textbf{89.93$\pm$0.31} & 88.77$\pm$0.65  & 89.49$\pm$0.40  & 88.24$\pm$0.55  & 85.01$\pm$1.42  & 89.60$\pm$1.05 \\
 & Kappa/AUCPR & \textbf{90.11$\pm$0.15} & 89.03$\pm$0.61  & 89.86$\pm$0.33  & 89.14$\pm$1.16  & 85.16$\pm$1.18  & 90.00$\pm$1.56 \\
\midrule

\multirow{3}{*}{\textbf{HMC}}
 & B-Acc       & \textbf{74.33$\pm$0.21} & 73.03$\pm$1.35  & 70.83$\pm$0.13  & 71.09$\pm$0.43  & 68.57$\pm$1.79  & 72.18$\pm$0.24 \\
 & F1/AUROC    & \textbf{77.31$\pm$0.20} & 76.26$\pm$1.62  & 74.24$\pm$0.21  & 71.86$\pm$0.79  & 69.14$\pm$1.16  & 74.16$\pm$0.27 \\
 & Kappa/AUCPR & \textbf{70.69$\pm$0.19} & 68.76$\pm$1.53  & 66.31$\pm$0.48  & 65.18$\pm$0.96  & 64.76$\pm$2.04  & 66.43$\pm$0.16 \\
\midrule

\multirow{3}{*}{\makecell{\textbf{Siena}}}
 & B-Acc       & \textbf{86.89$\pm$2.31} & 84.31$\pm$0.77  & 79.54$\pm$0.52  & 80.81$\pm$2.46  & 74.22$\pm$1.29  & 83.96$\pm$0.55 \\
 & F1/AUROC    & 89.12$\pm$2.30 & 90.30$\pm$1.94  & 87.63$\pm$2.61  & 88.07$\pm$2.56  & 82.90$\pm$2.17  & \textbf{93.15$\pm$1.49} \\
 & Kappa/AUCPR & \textbf{99.85$\pm$0.03} & 99.83$\pm$0.03  & 99.81$\pm$0.05  & 99.81$\pm$0.16  & 99.71$\pm$0.03  & 99.84$\pm$0.04 \\
\midrule

\multirow{3}{*}{\textbf{TUSL}}
 & B-Acc       & \textbf{79.82$\pm$1.95} & 78.33$\pm$1.59  & 33.33$\pm$0.00  & 61.30$\pm$2.93  & 62.44$\pm$1.75  & 77.54$\pm$1.95 \\
 & F1/AUROC    & \textbf{77.34$\pm$2.30} & 76.13$\pm$2.35  & 25.71$\pm$0.03  & 43.66$\pm$3.46  & 54.64$\pm$1.72  & 73.63$\pm$2.70 \\
 & Kappa/AUCPR & \textbf{65.45$\pm$2.31} & 64.60$\pm$2.86  & 00.00$\pm$0.00  & 31.30$\pm$2.01  & 41.21$\pm$1.57  & 65.12$\pm$3.48 \\
\midrule

\multirow{3}{*}{\textbf{Mimul-11}}
 & B-Acc       & \textbf{51.33$\pm$0.24} & 49.60$\pm$0.42  & 40.13$\pm$0.11  & 44.00$\pm$2.11  & 41.01$\pm$1.10  & 50.51$\pm$0.63 \\
 & F1/AUROC    & \textbf{58.82$\pm$0.23} & 55.23$\pm$0.55  & 46.77$\pm$0.14  & 48.92$\pm$1.91  & 44.59$\pm$0.97  & 58.09$\pm$0.61 \\
 & Kappa/AUCPR & 31.49$\pm$0.98 & 29.13$\pm$1.10  & 13.37$\pm$0.28  & 15.24$\pm$1.76  & 12.09$\pm$2.15  & \textbf{33.63$\pm$1.80} \\
\midrule

\multirow{3}{*}{\makecell{\textbf{Things}\\\textbf{EEG 2}}}
 & B-Acc       & \textbf{65.15$\pm$0.48} & 64.14$\pm$0.91  & 50.00$\pm$0.01  & 55.46$\pm$1.50  & 53.38$\pm$1.75  & 63.76$\pm$1.57 \\
 & F1/AUROC    & \textbf{86.63$\pm$0.35} & 75.02$\pm$0.84  & 59.97$\pm$0.10  & 61.48$\pm$0.46  & 57.56$\pm$1.66  & 77.54$\pm$0.70 \\
 & Kappa/AUCPR & 42.82$\pm$1.19 & \textbf{43.70$\pm$1.96}  & 15.78$\pm$0.02  & 17.21$\pm$0.34  & 15.08$\pm$1.14  & 40.89$\pm$1.20 \\
\midrule

\multirow{3}{*}{\textbf{SEED-V}}
 & B-Acc       & \textbf{47.40$\pm$0.25} & 42.27$\pm$1.52  & 22.06$\pm$0.12  & 29.58$\pm$0.92  & 26.30$\pm$1.51  & 40.98$\pm$1.58 \\
 & F1/AUROC    & \textbf{48.39$\pm$0.42} & 42.42$\pm$1.32  & 16.61$\pm$0.19  & 29.78$\pm$1.16  & 25.85$\pm$1.45  & 41.18$\pm$1.46 \\
 & Kappa/AUCPR & \textbf{34.84$\pm$0.45} & 27.59$\pm$1.33  & 2.56$\pm$0.25  & 12.01$\pm$1.09  & 8.62$\pm$1.71  & 25.58$\pm$1.87 \\
\midrule

\multirow{3}{*}{\textbf{ADFTD}}
 & B-Acc       & \textbf{58.75$\pm$0.66} & 56.43$\pm$0.95  & 49.82$\pm$0.23  & 51.58$\pm$1.56  & 49.46$\pm$2.40  & 52.20$\pm$0.93 \\
 & F1/AUROC    & \textbf{64.12$\pm$0.63} & 59.93$\pm$1.15  & 47.26$\pm$0.38  & 53.57$\pm$2.54  & 54.08$\pm$2.06  & 55.29$\pm$1.50 \\
 & Kappa/AUCPR & \textbf{39.41$\pm$0.79} & 37.07$\pm$1.36  & 28.96$\pm$0.29  & 28.71$\pm$2.51  & 28.41$\pm$1.48  & 31.38$\pm$1.82 \\

\bottomrule
\end{tabular}}
\end{table}

\subsection{Multi-task Comparison with Baselines}
\label{appx:multitask_baselines}
Table~\ref{tab:multitask_average} reports an additional full-parameter multi-task comparison with BENDR, BIOT, LaBraM, EEGPT, CBraMod, CSBrain, and REVE. All seven baselines use average-pooling classification heads.
This comparison assesses performance when the existing models also use joint training. PRiSE-EEG achieves Balanced Accuracy of 82.47\%, 65.55\%, 74.33\%, and 47.40\% on Workload, PhysioMI, HMC, and SEED-V, respectively; the highest baseline means are 74.67\%, 58.82\%, 72.18\%, and 43.26\%. But baselines still obtain higher means on TUSL, illustrating task-dependent differences.

\begin{table}[t]
\centering
\caption{Performance comparison of baseline models and PRiSE-EEG under full-parameter multi-task results on the 12 BCI tasks.}
\label{tab:multitask_average}
\setlength{\tabcolsep}{3pt}
\resizebox{\textwidth}{!}{
\begin{tabular}{llcccccccc}
\toprule
Dataset & Metric & BENDR & BIOT & LaBraM & EEGPT & CBraMod & CSBrain & REVE & PRiSE-EEG \\
\midrule
\multirow{3}{*}{SEED} & B-Acc & $60.82\pm0.19$ & $65.62\pm0.92$ & $67.71\pm0.55$ & $68.75\pm0.40$ & $72.25\pm1.90$ & $71.92\pm0.39$ & $72.62\pm0.92$ & $\mathbf{73.92\pm0.31}$ \\
& W-F1 & $60.84\pm0.20$ & $65.57\pm0.81$ & $68.11\pm0.66$ & $67.42\pm0.78$ & $72.90\pm1.63$ & $71.84\pm0.20$ & $72.75\pm0.76$ & $\mathbf{73.29\pm0.22}$ \\
& Kappa & $41.59\pm0.16$ & $50.11\pm0.99$ & $52.22\pm0.83$ & $52.37\pm0.28$ & $58.68\pm2.91$ & $58.02\pm0.59$ & $\mathbf{62.57\pm0.94}$ & $61.02\pm0.45$ \\
\midrule
\multirow{3}{*}{PhysioMI} & B-Acc & $44.77\pm0.31$ & $33.02\pm0.29$ & $43.19\pm1.03$ & $50.52\pm0.80$ & $41.80\pm0.69$ & $55.43\pm0.37$ & $58.82\pm0.45$ & $\mathbf{65.55\pm0.38}$ \\
& W-F1 & $44.63\pm0.03$ & $32.75\pm0.36$ & $42.38\pm1.15$ & $50.26\pm0.76$ & $40.35\pm0.87$ & $55.10\pm0.63$ & $58.77\pm0.56$ & $\mathbf{65.83\pm0.56}$ \\
& Kappa & $26.33\pm0.40$ & $10.67\pm0.40$ & $24.25\pm1.37$ & $34.00\pm1.06$ & $22.42\pm0.92$ & $40.48\pm0.48$ & $45.10\pm0.60$ & $\mathbf{54.78\pm0.69}$ \\
\midrule
\multirow{3}{*}{Workload} & B-Acc & $63.32\pm2.29$ & $71.37\pm2.19$ & $67.77\pm0.95$ & $72.66\pm0.90$ & $74.12\pm0.79$ & $71.27\pm0.75$ & $74.67\pm1.46$ & $\mathbf{82.47\pm0.30}$ \\
& AUROC & $74.09\pm1.04$ & $72.43\pm1.94$ & $77.32\pm2.88$ & $81.16\pm3.42$ & $83.89\pm0.93$ & $79.63\pm1.20$ & $82.32\pm1.47$ & $\mathbf{89.21\pm0.19}$ \\
& AUC-PR & $54.78\pm1.84$ & $69.52\pm0.23$ & $61.13\pm3.03$ & $64.30\pm6.48$ & $\mathbf{72.58\pm1.55}$ & $55.10\pm2.21$ & $68.22\pm3.92$ & $70.75\pm0.47$ \\
\midrule
\multirow{3}{*}{TUEV} & B-Acc & $67.46\pm2.59$ & $61.35\pm1.35$ & $71.53\pm0.19$ & $70.85\pm1.16$ & $69.41\pm1.86$ & $71.82\pm0.69$ & $74.27\pm0.87$ & $\mathbf{75.28\pm0.51}$ \\
& W-F1 & $84.56\pm1.82$ & $80.55\pm0.46$ & $86.22\pm0.73$ & $86.35\pm0.47$ & $83.44\pm0.31$ & $87.00\pm0.64$ & $\mathbf{89.37\pm0.33}$ & $88.25\pm0.51$ \\
& Kappa & $72.85\pm2.87$ & $68.37\pm0.61$ & $77.42\pm1.20$ & $77.69\pm1.06$ & $72.02\pm0.65$ & $77.84\pm1.55$ & $\mathbf{82.12\pm0.76}$ & $80.90\pm0.62$ \\
\midrule
\multirow{3}{*}{TUAB} & B-Acc & $81.72\pm0.26$ & $80.85\pm0.36$ & $80.47\pm0.12$ & $80.68\pm0.37$ & $81.55\pm0.14$ & $79.18\pm0.48$ & $81.18\pm0.46$ & $\mathbf{82.22\pm0.12}$ \\
& AUROC & $89.88\pm0.27$ & $88.67\pm0.62$ & $88.33\pm0.59$ & $\mathbf{90.10\pm0.52}$ & $89.57\pm0.26$ & $86.32\pm0.28$ & $88.32\pm0.37$ & $89.93\pm0.31$ \\
& AUC-PR & $90.22\pm0.15$ & $89.01\pm0.47$ & $88.65\pm0.28$ & $\mathbf{90.35\pm0.54}$ & $89.67\pm0.02$ & $85.78\pm0.25$ & $89.33\pm0.61$ & $90.11\pm0.15$ \\
\midrule
\multirow{3}{*}{HMC} & B-Acc & $70.21\pm0.72$ & $71.45\pm1.05$ & $71.63\pm1.04$ & $71.30\pm0.91$ & $72.03\pm0.14$ & $72.18\pm0.24$ & $71.93\pm0.45$ & $\mathbf{74.33\pm0.21}$ \\
& W-F1 & $72.44\pm0.65$ & $72.89\pm0.34$ & $72.28\pm1.08$ & $72.46\pm0.96$ & $75.20\pm0.29$ & $74.16\pm0.27$ & $72.07\pm0.73$ & $\mathbf{77.31\pm0.20}$ \\
& Kappa & $64.57\pm0.90$ & $64.89\pm0.71$ & $65.22\pm1.23$ & $64.86\pm0.77$ & $67.43\pm0.12$ & $66.43\pm0.16$ & $64.40\pm0.54$ & $\mathbf{70.69\pm0.19}$ \\
\midrule
\multirow{3}{*}{Siena} & B-Acc & $78.96\pm3.08$ & $74.35\pm0.71$ & $73.62\pm0.62$ & $83.28\pm1.56$ & $82.75\pm2.10$ & $76.42\pm1.43$ & $81.87\pm2.03$ & $\mathbf{86.89\pm2.31}$ \\
& AUROC & $91.31\pm4.74$ & $81.20\pm0.92$ & $89.43\pm3.59$ & $\mathbf{93.44\pm0.12}$ & $92.03\pm1.90$ & $92.70\pm0.67$ & $90.70\pm1.25$ & $89.12\pm2.30$ \\
& AUC-PR & $99.81\pm0.11$ & $99.26\pm0.05$ & $99.82\pm0.07$ & $99.86\pm0.00$ & $99.83\pm0.04$ & $99.87\pm0.05$ & $\mathbf{99.88\pm0.03}$ & $99.85\pm0.03$ \\
\midrule
\multirow{3}{*}{TUSL} & B-Acc & $73.35\pm0.98$ & $69.90\pm4.10$ & $75.86\pm1.15$ & $76.12\pm2.05$ & $79.56\pm3.24$ & $\mathbf{80.56\pm0.91}$ & $79.85\pm1.27$ & $79.82\pm1.95$ \\
& W-F1 & $70.90\pm1.73$ & $71.03\pm5.19$ & $72.74\pm1.33$ & $\mathbf{84.19\pm1.04}$ & $76.24\pm4.45$ & $76.72\pm0.38$ & $77.81\pm1.89$ & $77.34\pm2.30$ \\
& Kappa & $56.65\pm2.55$ & $56.00\pm7.50$ & $59.60\pm1.81$ & $\mathbf{84.30\pm1.27}$ & $64.72\pm5.96$ & $65.57\pm0.41$ & $66.62\pm2.35$ & $65.45\pm2.31$ \\
\midrule
\multirow{3}{*}{Mimul-11} & B-Acc & $50.98\pm1.52$ & $40.43\pm0.63$ & $45.05\pm0.66$ & $50.77\pm0.61$ & $44.53\pm0.19$ & $41.64\pm0.54$ & $45.22\pm0.13$ & $\mathbf{51.33\pm0.24}$ \\
& W-F1 & $57.85\pm2.80$ & $49.02\pm0.73$ & $53.42\pm1.98$ & $57.44\pm1.11$ & $52.80\pm0.57$ & $50.70\pm0.61$ & $53.73\pm0.65$ & $\mathbf{58.82\pm0.23}$ \\
& Kappa & $\mathbf{36.72\pm4.93}$ & $14.63\pm1.88$ & $22.60\pm2.20$ & $31.16\pm2.43$ & $19.23\pm0.77$ & $15.94\pm1.33$ & $21.53\pm1.16$ & $31.49\pm0.98$ \\
\midrule
\multirow{3}{*}{Things-EEG-2} & B-Acc & $63.62\pm1.38$ & $54.15\pm0.25$ & $56.47\pm0.32$ & $51.54\pm0.76$ & $59.13\pm0.72$ & $57.49\pm0.27$ & $59.43\pm1.20$ & $\mathbf{65.15\pm0.48}$ \\
& AUROC & $77.23\pm0.88$ & $63.01\pm0.52$ & $62.82\pm0.84$ & $66.64\pm0.04$ & $69.58\pm1.41$ & $62.47\pm1.27$ & $69.00\pm0.53$ & $\mathbf{86.63\pm0.35}$ \\
& AUC-PR & $42.29\pm1.75$ & $20.12\pm0.33$ & $22.23\pm0.64$ & $20.30\pm0.65$ & $31.85\pm1.21$ & $24.66\pm1.93$ & $32.07\pm1.85$ & $\mathbf{42.82\pm1.19}$ \\
\midrule
\multirow{3}{*}{SEED-V} & B-Acc & $21.75\pm0.66$ & $30.54\pm0.18$ & $43.26\pm0.29$ & $43.07\pm0.60$ & $40.56\pm0.77$ & $38.02\pm0.23$ & $40.84\pm0.98$ & $\mathbf{47.40\pm0.25}$ \\
& W-F1 & $19.94\pm2.58$ & $30.97\pm0.26$ & $43.20\pm0.81$ & $42.99\pm0.54$ & $41.72\pm0.72$ & $38.07\pm1.03$ & $40.58\pm2.00$ & $\mathbf{48.39\pm0.42}$ \\
& Kappa & $2.90\pm0.68$ & $14.14\pm0.21$ & $28.93\pm0.50$ & $28.62\pm0.86$ & $26.22\pm0.97$ & $22.48\pm0.42$ & $25.59\pm1.48$ & $\mathbf{34.84\pm0.45}$ \\
\midrule
\multirow{3}{*}{ADFTD} & B-Acc & $55.71\pm3.76$ & $52.40\pm1.98$ & $50.27\pm1.74$ & $52.91\pm1.57$ & $51.95\pm2.84$ & $48.96\pm2.65$ & $52.12\pm3.78$ & $\mathbf{58.75\pm0.66}$ \\
& W-F1 & $56.97\pm2.75$ & $53.48\pm2.71$ & $52.36\pm2.23$ & $54.82\pm1.30$ & $54.41\pm3.13$ & $50.17\pm3.76$ & $51.10\pm3.90$ & $\mathbf{64.12\pm0.63}$ \\
& Kappa & $33.98\pm4.63$ & $36.27\pm3.65$ & $26.17\pm3.15$ & $31.14\pm1.85$ & $30.06\pm4.35$ & $24.96\pm4.13$ & $30.70\pm6.41$ & $\mathbf{39.41\pm0.79}$ \\
\bottomrule
\end{tabular}}
\end{table}

\subsection{Additional Ablation Results}\label{appx:extra_ablation}

\begin{table}[t]
\centering
\caption{Balanced Accuracy for the full model and ablation variants (mean $\pm$ standard deviation, \%).}
\label{tab:ablation_numeric}
\resizebox{\linewidth}{!}{
\begin{tabular}{lcccc}
\toprule
Configuration & TUSL & SEED-V & Workload & HMC \\
\midrule
Full PRiSE-EEG & $79.82\pm1.95$ & $47.40\pm0.25$ & $82.47\pm0.30$ & $74.33\pm0.21$ \\
Static path only & $67.31\pm1.65$ & $43.41\pm0.20$ & $78.15\pm0.49$ & $71.13\pm0.59$ \\
Dynamic path only & $64.89\pm1.77$ & $40.75\pm0.86$ & $74.93\pm1.92$ & $72.59\pm0.30$ \\
Without Gumbel selection & $69.11\pm2.02$ & $46.31\pm0.44$ & $78.89\pm0.43$ & $72.24\pm0.39$ \\
Two additional shared experts per layer & $77.12\pm1.40$ & $46.12\pm0.60$ & $80.11\pm0.67$ & $72.01\pm0.69$ \\
Standard dense FFN & $73.88\pm0.94$ & $45.19\pm0.45$ & $79.25\pm0.60$ & $71.40\pm0.42$ \\
Reversed expert allocation & $58.80\pm2.72$ & $29.58\pm1.99$ & $65.26\pm1.36$ & $66.40\pm1.14$ \\
Flattened classification head & $77.92\pm1.01$ & $39.41\pm0.49$ & $79.56\pm0.73$ & $72.70\pm0.30$ \\
\bottomrule
\end{tabular}}
\end{table}

Table~\ref{tab:ablation_numeric} lists the detailed numbers for Balanced Accuracy of the ablation variants shown in Fig.~\ref{fig:ablation}. The fixed allocation uses two additional shared experts from the eight-expert pool, giving three shared and six specialized experts in total per layer.
Using the full-model values as the reference, combining static and dynamic paths yields higher Balanced Accuracy than either path alone on all four tasks. Relative to this fixed allocation, the depth-stratified configuration improves TUSL, SEED-V, Workload, and HMC by 2.70\%, 1.28\%, 2.36\%, and 2.32\%, respectively. Reversing the allocation lowers the corresponding results by 21.02\%, 17.82\%, 17.21\%, and 7.93\%.

We first evaluate the sensitivity of the attention bias strength (ranging from 0 to $-10^{6}$) within the Prior-Guided Continuous Tokenization, as illustrated in Fig.~\ref{fig:ablation-prior-mask}.
Compared to the baseline without prior guidance (bias = 0), the application of a soft attention bias yields measurable performance gains of 5-10\% across the TUSL, Workload, and SEED-V datasets.
However, as the bias strength approaches a hard mask configuration, performance across all three datasets progressively declines.
Furthermore, enforcing a hard mask (bias = $-\infty$) induces training instability during pre-training, as the model fails to optimize the parameters in the layers preceding the tokenizer.
This phenomenon is likely attributable to an information bottleneck generated by the sparse attention mask, which obstructs effective gradient propagation.
These findings validate the tokenizer design, confirming that a soft, non-zero attention bias is optimal for learning versatile EEG representations.
We also investigate the optimal stage for introducing neuroscientific priors.
As shown in Fig.~\ref{fig:ablation-prior-stage}, incorporating prior knowledge during pre-training results in substantially better downstream performance compared to its introduction during the fine-tuning stage alone.
This result underscores the importance of anchoring the model to neuroscientific principles from the onset of the learning process.

\subsection{Prior Preservation and Grouping Sanity Check}
\label{appx:prior_preservation}

PRiSE-EEG uses static region-network priors only as weak sensor-space biases, and the learnable relaxation term $\Delta_{\mathrm{pri}}$ is expected to adjust rather than overwrite the initial grouping structure.
We therefore measure whether the optimized prior matrix $\mathbf{M}_s+\Delta_{\mathrm{pri}}$ preserves the coarse in-group/out-group separation after training.
In Tab.~\ref{tab:prior_preservation}, we report the relative update magnitude $\|\Delta_{\mathrm{pri}}\|/\|\mathbf{M}_s\|$, the average in-group bias, the average out-group bias, and the in-group minus out-group gap.
We also compare against a random learnable grouping baseline under the same training budget.

\begin{table}[H]
\centering
\caption{
Prior preservation and grouping sanity check.
The full PRiSE-EEG prior keeps a clear gap after training, while random learnable grouping shows a weaker structural separation.
}
\label{tab:prior_preservation}
\resizebox{\linewidth}{!}{
\begin{tabular}{lcccc}
\toprule
Variant & $\|\Delta_{\mathrm{pri}}\|/\|\mathbf{M}_s\|$ $\downarrow$ & In-group bias $\uparrow$ & Out-group bias $\downarrow$ & In-group $-$ out-group $\uparrow$ \\
\midrule
Full PRiSE-EEG prior & $0.2801{\pm}0.0517$ & $-0.1638{\pm}0.0405$ & $-1.0487{\pm}0.0310$ & $0.8849$ \\
Random learnable grouping & $0.4345{\pm}0.0892$ & $-0.6911{\pm}0.0783$ & $-0.8255{\pm}0.0712$ & $0.1344$ \\
\bottomrule
\end{tabular}}
\end{table}

The learned update remains modest relative to the initial prior matrix, and the in-group/out-group gap is largely preserved.
This indicates that the prior relaxation term provides data-driven correction without collapsing the predefined sensor-group structure.

\subsection{Comparison with Recent EEG Tokenizers}
\label{appx:tokenizer_comparison}

MMM organizes spatial information using a unified topology and a multi-level channel hierarchy, with regional representations supporting different channel configurations focusing on emotion recognition~\citep{yi2023mmm}. PRiSE-EEG shares the idea of structured spatial aggregation, but turns static region-network groupings into learnable soft attention biases and adds sample-dependent short-time dynamic groups. Soft biases allow out-of-group channels to participate, while dynamic groups supplement sample- and time-varying channel relations. Together they form continuous multi-channel tokens; the subsequent CKA-calibrated experts organize shared and specialized capacity across depth.

We compare PRiSE-EEG with recent tokenizer-centered EEG foundation models.
CodeBrain uses a decoupled temporal-frequency discrete tokenizer with a multi-scale EEGSSM backbone, while TFM-Tokenizer learns single-channel time-frequency motif tokens.
In contrast, PRiSE-EEG keeps EEG patches continuous and multi-channel, and combines weak prior-guided tokenization with CKA-calibrated depth-stratified experts.
PRiSE-EEG outperforms both discrete-tokenizer baselines on the shared evaluation datasets under the matched pipeline.
These results suggest that continuous multi-channel prior-guided tokenization is complementary to discrete EEG vocabularies, and that the proposed depth-stratified expert allocation provides additional cross-paradigm capacity.

\begin{table}[H]
\centering
\small
\caption{
Comparison with recent tokenizer-centered EEG foundation models.
``Reported'' follows the original paper when available; ``released code'' follows the public implementation; ``our pipeline'' uses the same preprocessing, split cache, seed control, and fine-tuning budget as PRiSE-EEG.
Balanced Accuracy is reported.
}
\label{tab:tokenizer_comparison}
\resizebox{0.8\linewidth}{!}{
\begin{tabular}{lcccc}
\toprule
Model & TUEV & SEED-V & HMC & Workload \\
\midrule
CodeBrain reported & $64.28{\pm}0.62$ & $41.37{\pm}0.23$ & -- & $75.14{\pm}2.03$ \\
CodeBrain by released code & $63.19{\pm}1.06$ & $41.24{\pm}0.39$ & $71.93{\pm}0.54$ & $75.25{\pm}0.33$ \\
CodeBrain by our pipeline & $64.43{\pm}1.22$ & $41.54{\pm}0.45$ & $72.06{\pm}0.48$ & $78.73{\pm}0.32$ \\
TFM-Tokenizer reported & $49.43{\pm}5.16$ & -- & -- & -- \\
TFM-Tokenizer by released code & $51.12{\pm}0.60$ & $35.50{\pm}0.18$ & $66.74{\pm}0.54$ & $58.47{\pm}0.59$ \\
TFM-Tokenizer by our pipeline & $53.65{\pm}0.55$ & $36.31{\pm}0.28$ & $66.10{\pm}0.43$ & $60.35{\pm}0.52$ \\
\textbf{PRiSE-EEG} & $\textbf{75.25}{\pm}0.48$ & $\textbf{47.34}{\pm}0.23$ & $\textbf{74.30}{\pm}0.24$ & $\textbf{82.44}{\pm}0.28$ \\
\bottomrule
\end{tabular}}
\end{table}

\section{Visualization}
\label{appx:visual}

\subsection{Interpretable Decision Localization}

\begin{figure}[b]
\centering
\includegraphics[width=0.6\linewidth]{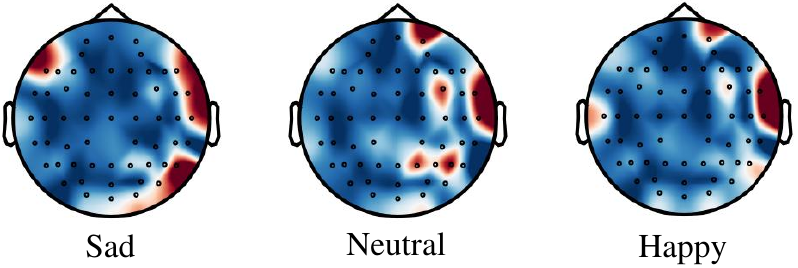}
\caption{Grad-CAM visualization of PRiSE-EEG on SEED showing the region of model interest for Sad, Neutral and Happy emotion. Red colors indicating high relevance.}
\label{fig:vis-topomap}
\end{figure}

\begin{figure}[t]
  \centering
  \includegraphics[width=0.6\linewidth]{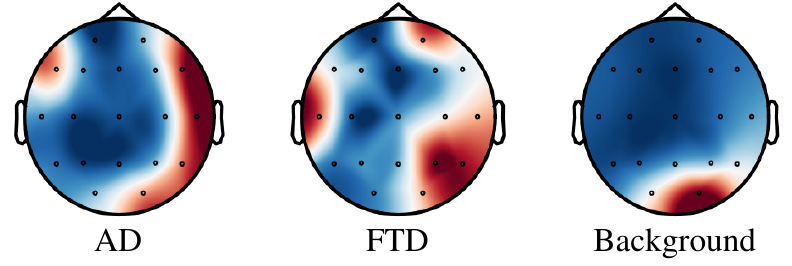}
  \caption{Grad-CAM visualization of our PRiSE-EEG on ADFTD dataset showing the region of model interest for Alzheimer's Disease, Frontotemporal Dementia and healthy control group. Red colors indicating high relevance.}
  \label{fig:grad-cam-adftd}
\end{figure}

\begin{figure}[t]
  \centering
  \begin{subfigure}[t]{0.22\textwidth}
    \includegraphics[width=\linewidth]{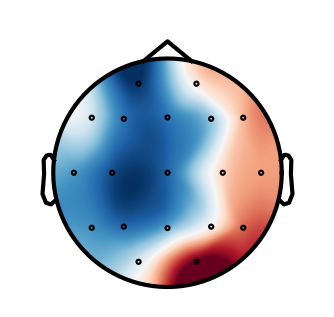}
    \caption{Arithmetic}
    \label{fig:grad-cam-workload-arithmetic}
  \end{subfigure}
  \begin{subfigure}[t]{0.22\textwidth}
    \includegraphics[width=\linewidth]{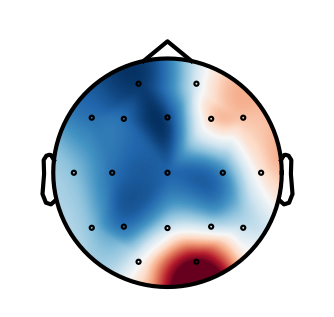}
    \caption{Background}
    \label{fig:grad-cam-workload-background}
  \end{subfigure}
  \caption{Grad-CAM visualization of PRiSE-EEG on Workload dataset showing the region of model interest for Arithmetic and Background status. Red colors indicating high relevance.}
  \label{fig:grad-cam-workload}
\end{figure}

\begin{figure}[t]
  \centering
  \begin{subfigure}[t]{0.22\textwidth}
    \includegraphics[width=\linewidth]{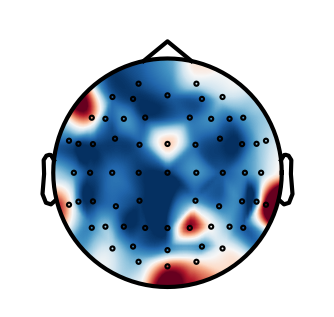}
    \caption{Target}
    \label{fig:grad-cam-things-eeg-2-target}
  \end{subfigure}
  \begin{subfigure}[t]{0.22\textwidth}
    \includegraphics[width=\linewidth]{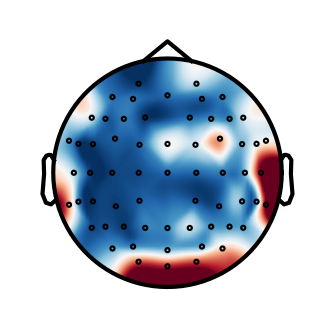}
    \caption{Non-Target}
    \label{fig:grad-cam-things-eeg-2-non-target}
  \end{subfigure}
  \caption{Grad-CAM visualization of PRiSE-EEG on Things-EEG-2 dataset showing the region of model interest for Target and Non-Target status. Red colors indicating high relevance.}
  \label{fig:grad-cam-things-eeg-2}
\end{figure}

We apply Grad-CAM~\citep{selvaraju2017grad} to the SEED, ADFTD, Workload, and Things-EEG-2 datasets to analyze how \textbf{PRiSE-EEG} extracts task-relevant spatial information from EEG signals.
Specifically, Grad-CAM is a visualization technique that highlights critical regions in the input data that influence the predictions of model.
By computing gradient-based weights from the target class score to the original EEG features, the resulting heatmaps provide intuitive explanations for model decisions.
Although class-specific attribution patterns may vary across individuals, the averaged Grad-CAM maps summarize the scalp regions that contribute to the model's predictions.
In our implementation, we generate heatmaps from the final layer of our dual-stream MLP classifier, allowing us to visualize channel-wise evidence for the model's decisions.
As shown in Fig.~\ref{fig:vis-topomap}, apparent lateralization is consistent with previous research on the functional networks of emotion~\citep{liu2023fine}.
For Sad emotion, our framework pays more attention to F7, F8, and P8 electrodes, suggesting that the right frontal and temporal lobes may be dominant in negative emotion.
For Neutral emotion, electrodes P2 and P6 receive relatively high attribution, indicating that the model uses parietal scalp features for this class. This spatial pattern may indicate activity in the Default Mode Network~(DMN).

As shown in Fig.~\ref{fig:grad-cam-adftd}, the Grad-CAM visualizations for ADFTD show different scalp attribution patterns for Alzheimer's Disease~(AD) and Frontotemporal Dementia~(FTD).
For the classification of AD, PRiSE-EEG assigns the highest importance to electrodes situated over the temporal and parietal lobes. 
When classifying FTD, attribution is more prominent at frontal electrodes.
This spatial distinction is broadly consistent with reported regional differences between AD and FTD~\citep{nardone2018ftd-ad}.
AD is characterized by pathology that predominantly affects posterior cortical areas, leading to classic EEG findings such as decreased alpha-band coherence in the temporal-parietal-occipital regions and general slowing of the background rhythm. In contrast, FTD is defined by progressive degeneration of the frontal and anterior temporal lobes, which can manifest as focal slowing.

As shown in Fig.~\ref{fig:grad-cam-workload}, the Workload visualizations highlight electrodes Fp2, F7, F8, C4, T8, P8, O2, and F4, with different spatial distributions between background and arithmetic conditions. In the non-math condition, attribution is concentrated at frontal electrodes Fp2, F7, and F8. During arithmetic, it extends to temporal and parietal sites, suggesting that the classifier draws on more distributed spatial evidence. This shift is broadly compatible with the involvement of frontal and parietal regions in mathematical processing~\citep{vansteensel2014spatiotemporal,artemenko2021fronto-partietal}. The contrast suggests that PRiSE-EEG uses condition-dependent spatial information to distinguish arithmetic workload from background recordings.

For Things-EEG-2, the Grad-CAM maps in Fig.~\ref{fig:grad-cam-things-eeg-2} show concentrated attribution over parietal and left frontal electrodes for target stimuli, whereas this pattern is less prominent for non-target stimuli. The parietal emphasis is compatible with the scalp distribution of target-related P300 responses, particularly the P3b component~\citep{boucher2010behavioural-p3b,polich2007updating-p300}. P3b is a positive-going ERP component that typically peaks approximately 300-600\,ms after a rare, task-relevant target and is most prominent over centro-parietal and parieto-occipital sites. It has been associated with attentional allocation, working-memory updating, and stimulus evaluation. These properties provide a possible physiological context for the target-related attribution pattern. The maps suggest that PRiSE-EEG uses spatial features relevant to target discrimination, although establishing that these features specifically reflect P3b would require time-resolved waveform analysis.

\subsection{Latent Feature Clustering via t-SNE}

\begin{figure*}[!ht]
    \centering
    \begin{subfigure}[t]{0.445\textwidth} 
        \includegraphics[width=\linewidth]{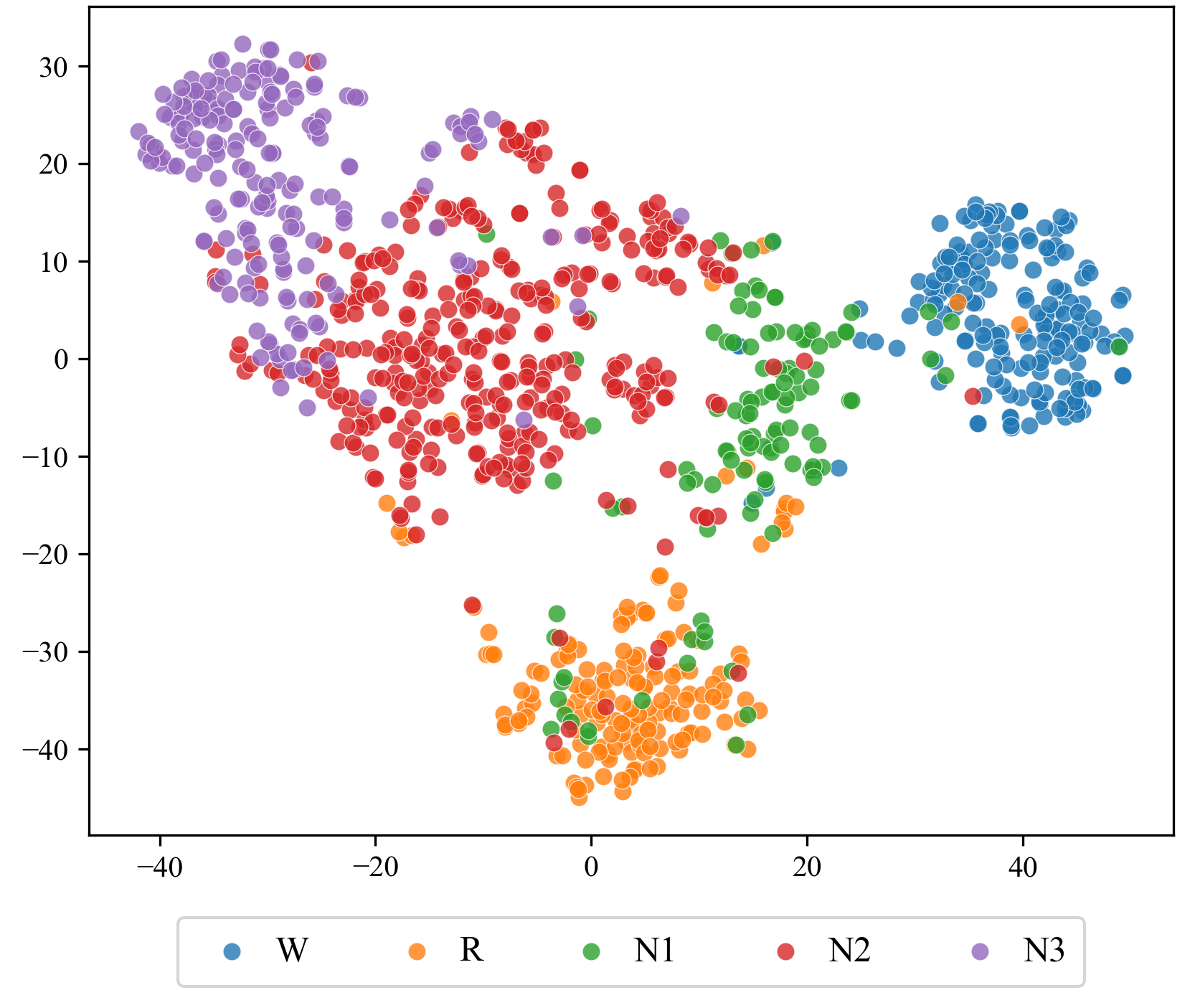}
        \caption{HMC}
        \label{fig:t-sne-hmc}
    \end{subfigure}
    \begin{subfigure}[t]{0.445\textwidth}
        \includegraphics[width=\linewidth]{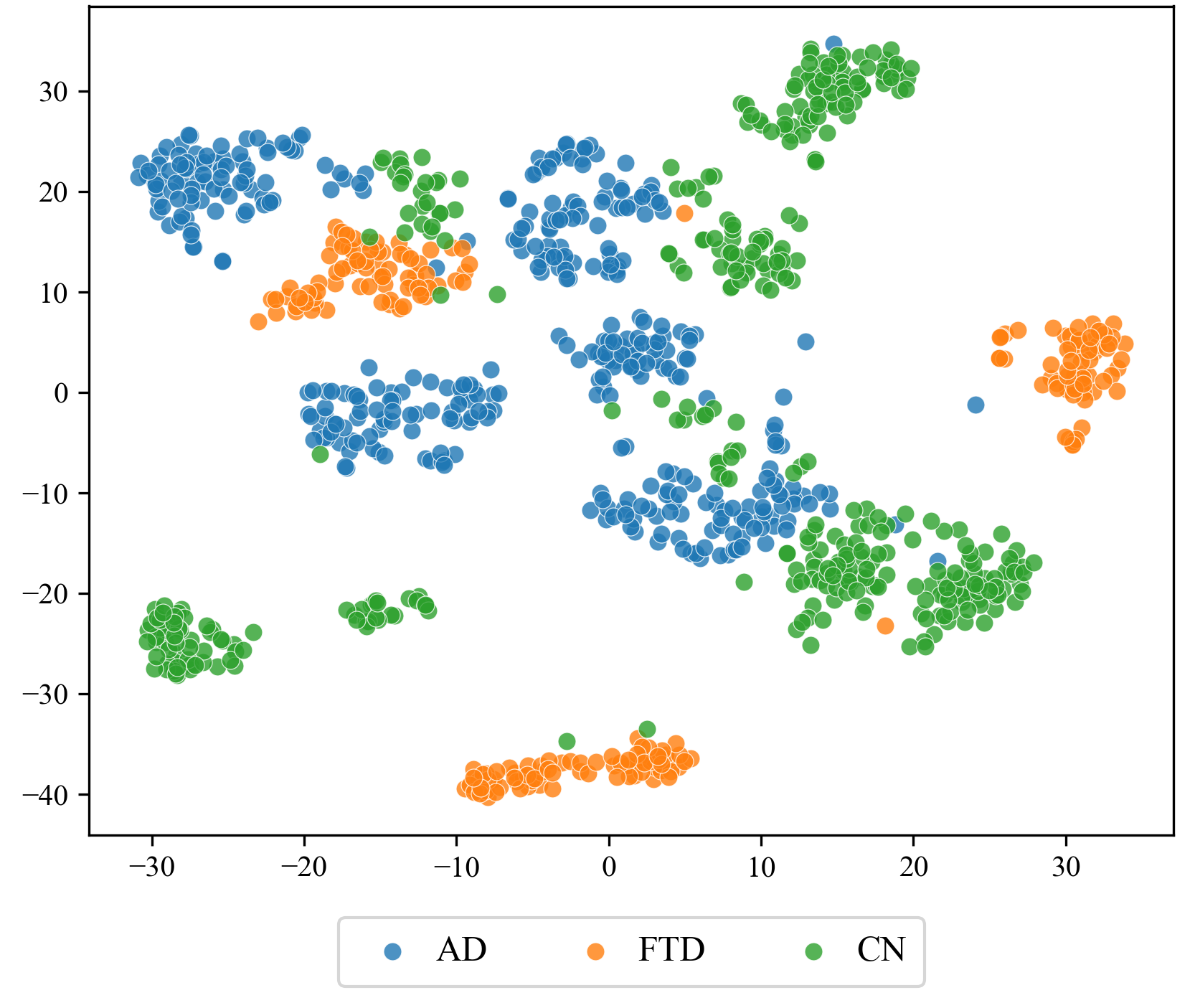}
        \caption{ADFTD}
        \label{fig:t-sne-adftd}
    \end{subfigure}

    \vspace{1em} 

    \begin{subfigure}[t]{0.445\textwidth}
        \includegraphics[width=\linewidth]{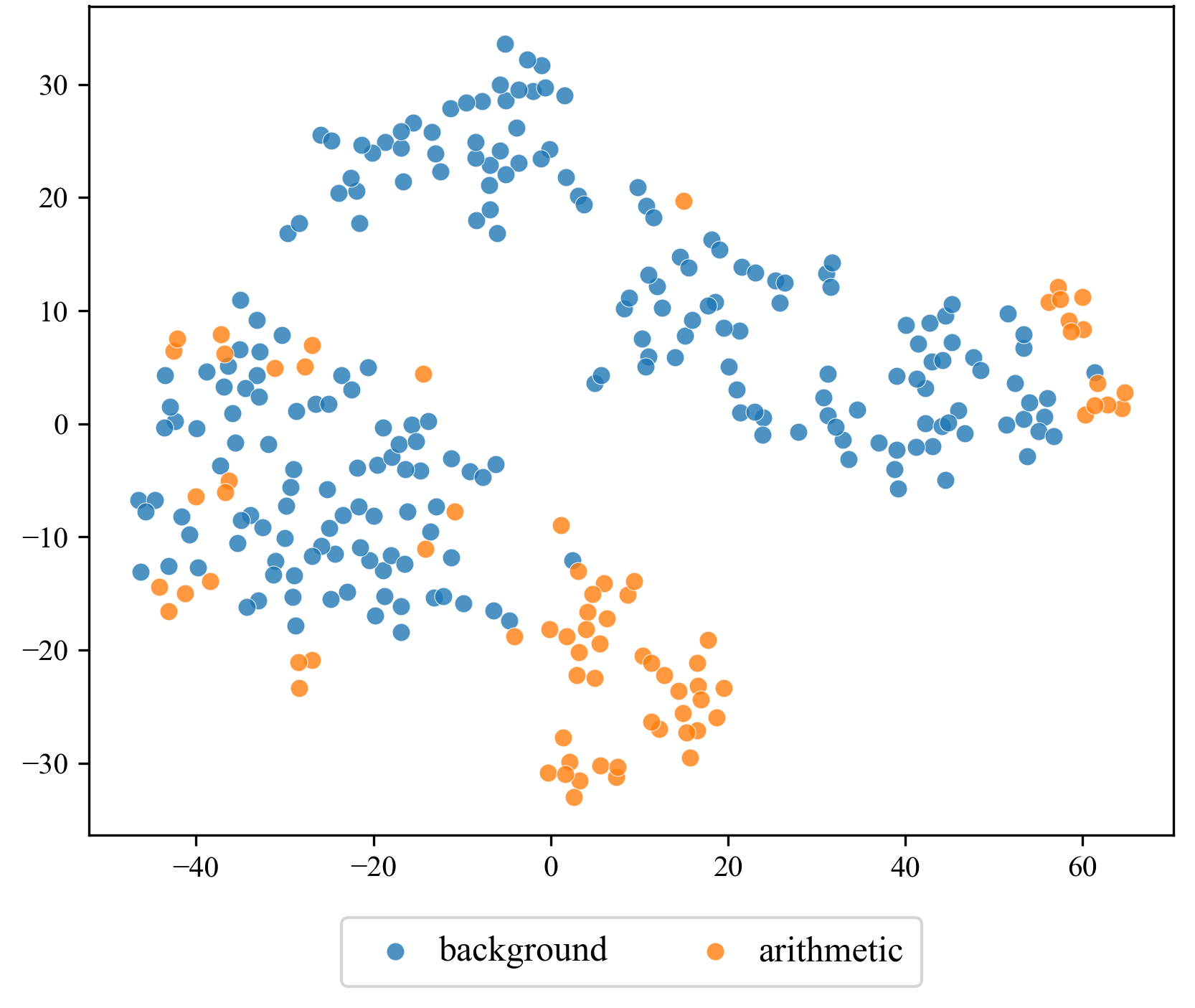}
        \caption{Workload}
        \label{fig:t-sne-workload}
    \end{subfigure}
    \begin{subfigure}[t]{0.445\textwidth}
        \includegraphics[width=\linewidth]{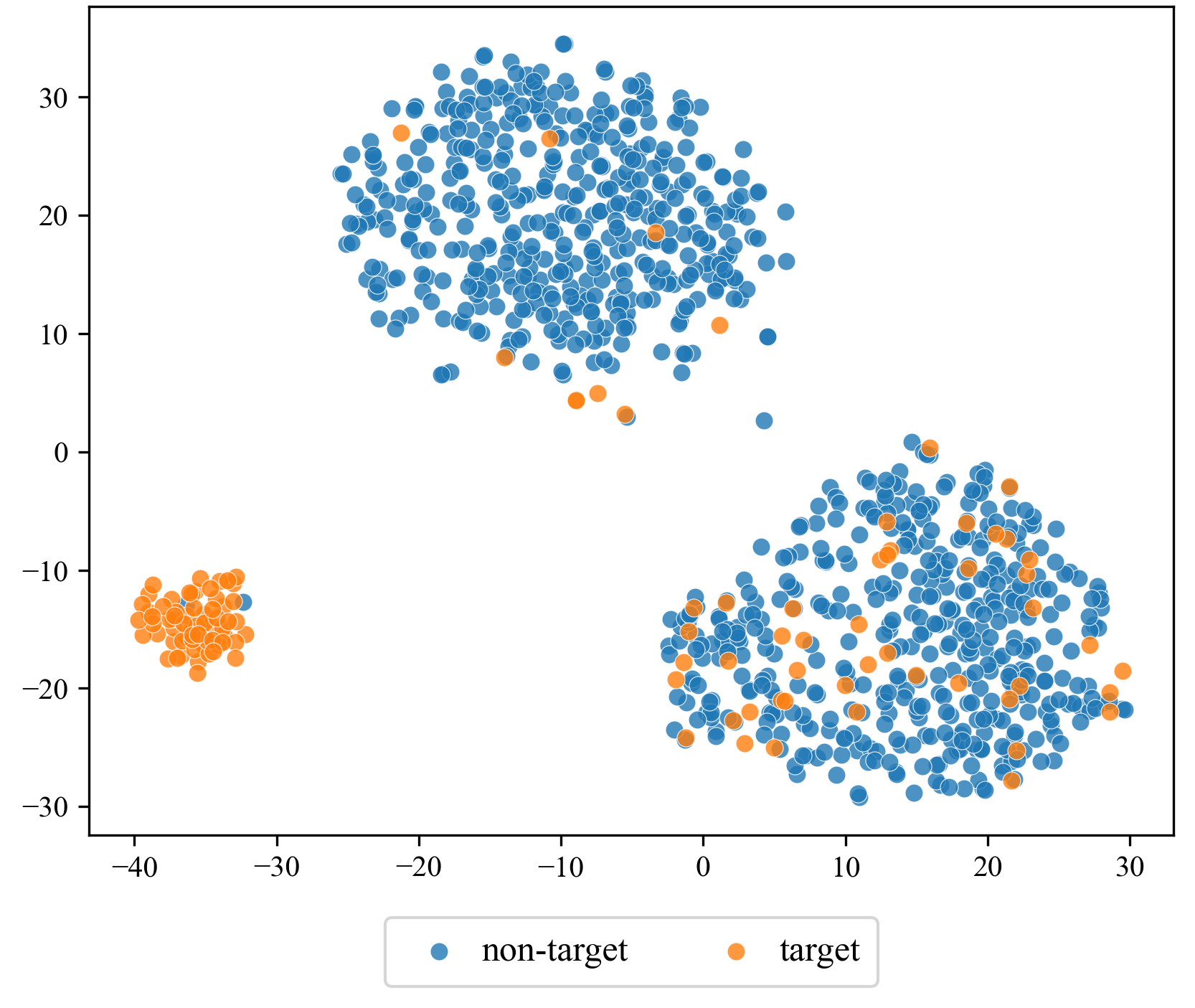}
        \caption{Things-EEG-2}
        \label{fig:t-sne-things-eeg-2}
    \end{subfigure}
    \caption{t-SNE visualizations of feature embeddings reduced to 40 dimensions by PCA on different datasets with PRiSE-EEG. t-SNE runs for 1000 iterations in which the perplexity is 30 and the number of samples is 100.}
    \label{fig:t-sne}
\end{figure*}

We use t-SNE~\citep{vandermaaten08a}, a nonlinear dimensionality reduction technique, to visualize the final classification features from four key datasets (Fig.~\ref{fig:t-sne}). This allows us to inspect the separability of features and demonstrate the representation ability of our framework.
\begin{itemize}
    \item For the HMC dataset, as shown in Fig.~\ref{fig:t-sne-hmc}, the five sleep stages form tightly grouped clusters, reflecting high intra-class consistency with physiological patterns, which also confirms the representation ability of PRiSE-EEG to disentangle discriminative features while tolerating noise.
    \item For the ADFTD dataset, the embeddings for AD, FTD, and healthy controls form separable clusters, corroborating the Grad-CAM findings as in Fig.~\ref{fig:t-sne-adftd}. These clusters are less compact than those for sleep staging, which accurately reflects the higher inter-subject variability and more subtle neurophysiological differences inherent in complex neurodegenerative diseases. Nonetheless, the model successfully carves out a feature space that captures these nuanced distinctions.
    \item For the Workload dataset, the two classes, ``arithmetic'' and ``background'' form two separated clusters in Fig.~\ref{fig:t-sne-workload}. This confirms that the model learns a robust representation that can reliably distinguish between a task-engaged cognitive state and a resting state, mirroring the functional network separation identified in the Grad-CAM analysis.
    \item For Things-EEG-2, the ``target and ``non-target classes form distinct clusters in the t-SNE visualization, suggesting that the learned representations contain information useful for target discrimination.
\end{itemize}

\section{Scaling Behavior}
\label{appx:scaling}

\begin{figure*}[ht]
  \centering
  \includegraphics[width=\linewidth]{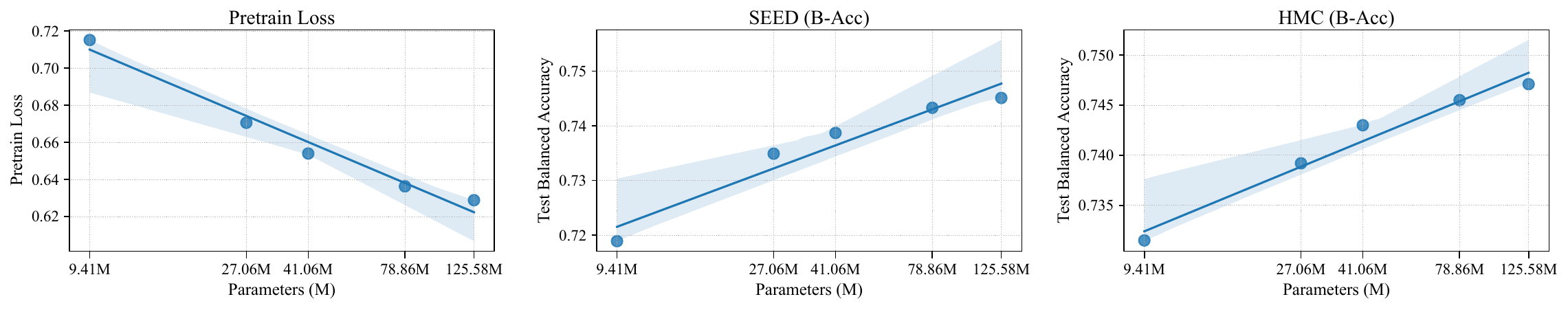}
  \caption{Scaling behavior with model size $N$: pretraining validation loss, SEED Balanced Accuracy, and HMC Balanced Accuracy.}
  \label{fig:scaling_param}
\end{figure*}

\begin{figure*}[ht]
  \centering
  \includegraphics[width=\linewidth]{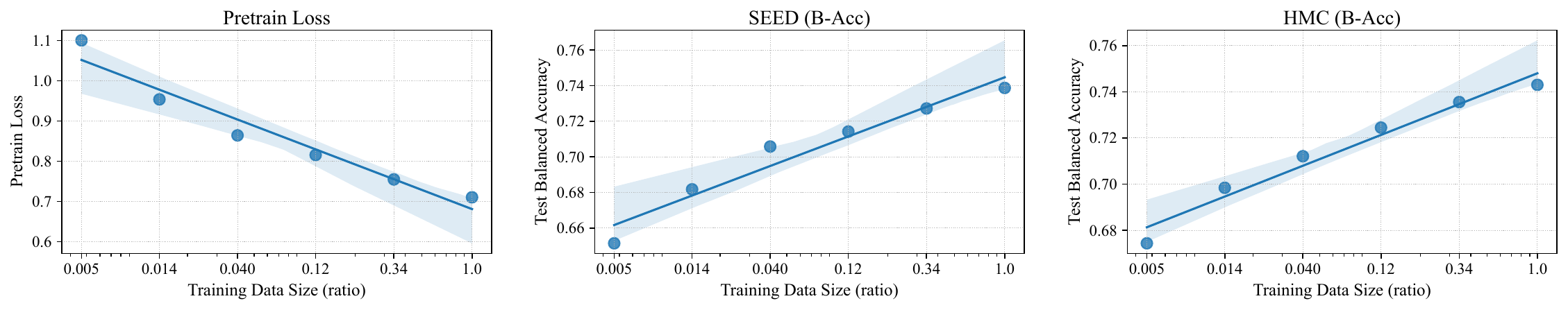}
  \caption{Scaling behavior with pretraining data size, ranging from 140 to 28,000 hours: pretraining validation loss, SEED Balanced Accuracy, and HMC Balanced Accuracy. The data ratio $P$ denotes the fraction of the full 28,000-hour corpus.}
  \label{fig:scaling_data}
\end{figure*}

In this section, we study the relationship between model performance and training scale, including the model size and pre-training data size.
To study the effect of model size, we use five size variants of PRiSE-EEG (Fig.~\ref{fig:scaling_param}) and track the impact on both pre-training loss and downstream performance.
To study the effect of data size, we vary the pre-training corpus from 140 to 28,000 hours and again evaluate the impact on pre-training loss and downstream performance.
Pre-training loss is measured on a held-out validation set of the pre-training data, while downstream Balanced Accuracy is reported separately for HMC and SEED. The data ratio $P$ is the number of pretraining hours divided by 28,000.

\subsection{For Model Size}
As shown in Fig.~\ref{fig:scaling_param}, our scaling law experiments reveal a clear and predictable relationship between model size and performance.
The results on the pre-training loss function show that the scaling law of the test $\mathcal{L}_p$ with model size~($N$) is: $\mathcal{L}_p$ = $-0.034$ * $\ln$(N) + $0.786$ , where $R^2$ is $0.974$.
The results on the SEED dataset show that the scaling law of the test balanced accuracy with model size~($N$) is: $BAcc$ = $0.010$ * $\ln$(N) + $0.699$, where $R^2$ is $0.940$.
The results on the HMC dataset show that the scaling law of the test balanced accuracy with model size~($N$) is: $BAcc$ = $0.006$ * $\ln$(N) + $0.719$, where $R^2$ is $0.969$.
The consistently high $R^2$ values across both pre-training and downstream tasks confirm a strong, predictable scaling behavior, which demonstrate that the PRiSE-EEG is robustly scalable and will reliably improve with increasing parameters.

\subsection{For Data Size}

In Fig.~\ref{fig:scaling_data}, we provide the results of the scale law experiments on the pre-training loss function~($\mathcal{L}_p$), SEED, and HMC dataset.
The results on the pre-training loss function show that the scaling law of the test $\mathcal{L}_p$ with data ratio~($P$) is: $\mathcal{L}_p$ = $-0.069$ * $\ln$(P) + $0.681$, where $R^2$ is $0.945$.
The results on the SEED dataset show that the scaling law of the test balanced accuracy with data ratio~($P$) is: $BAcc$ = $0.015$ * $\ln$(P) + $0.745$, where $R^2$ is $0.945$.
The results on the HMC dataset show that the scaling law of the test balanced accuracy with data ratio~($P$) is: $BAcc$ = $0.012$ * $\ln$(P) + $0.748$, where $R^2$ is $0.964$.
The high $R^2$ values across all metrics confirm that PRiSE-EEG's performance scales predictably and effectively with the volume of pre-training data. 
This indicates that performance is not yet saturated and will likely continue to improve as more data becomes available.


\section{Computational Cost and Analysis Overhead}
\label{appx:compute}
\subsection{Capacity-matched Computational Cost}
We use the PRiSE-EEG-B configuration for all cost measurements. Compared with the standard 41.06M-parameter model, the downstream instances include a measurement-specific four-class classification head, increasing the total parameter count to approximately 43.8M. We compare the depth-stratified MoE, a uniform two-shared-expert MoE, and a total-parameter-matched dense model.

Measurements use an RTX 4070 Ti SUPER, PyTorch 2.7.1, CUDA 12.8, BF16 autocast, FP32 parameters and AdamW states, eager execution and four CPU threads. Newly initialized models receive GPU-resident synthetic signals with 20 channels, 10 seconds, and 256\,Hz sampling. For every instance, inference uses 10 warm-up iterations and 40 timed iterations; training uses three warm-up iterations and 15 timed iterations. CUDA synchronization is performed for each measurement, and wall-clock means are reported.

An operator counter records supported matrix multiplication, convolution, and scaled dot-product attention operations, counting a multiply-add as two operations. FFT, normalization, activations, softmax, indexing, and optimizer arithmetic are excluded. Memory is PyTorch peak allocated memory. A complete fine-tuning step includes gradient clearing, forward evaluation, loss computation, backward propagation, gradient scaling and clipping, and an AdamW update; it excludes data loading, validation, checkpoint saving, and multi-GPU communication.

\begin{table}[ht]
\centering
\caption{Capacity-matched inference and fine-tuning costs. FLOPs are counted at $B=1$; memory and fine-tuning steps are measured at $B=8$.}
\label{tab:cost_inference}
\resizebox{\textwidth}{!}{
\begin{tabular}{lrrrrrrrr}
\toprule
Instance & \makecell{Total\\params (M)} & \makecell{Step\\GFLOPs/sample} & \makecell{$B=1$\\latency (ms)} & \makecell{$B=8$\\latency (ms)} & \makecell{$B=8$\\samples/s} & \makecell{Inference\\peak (MiB)} & \makecell{Fine-tuning\\step (ms)} & \makecell{Fine-tuning\\peak (MiB)} \\
\midrule
Stratified MoE & 43.790 & 20.457 & 17.37 & 28.09 & 284.78 & 414.1 & 77.67 & 2500.4 \\
Two-shared MoE & 43.793 & 18.972 & 17.66 & 28.81 & 277.67 & 413.6 & 80.70 & 2419.8 \\
Total-parameter-matched dense & 43.761 & 32.514 & 5.67 & 18.02 & 443.94 & 460.3 & 52.44 & 2339.8 \\
\bottomrule
\end{tabular}}
\end{table}

Stratified and uniform sharing have similar measured inference latencies: 17.37 versus 17.66\,ms at $B=1$, and 28.09 versus 28.81\,ms at $B=8$, corresponding to 284.78 and 277.67 samples/s. Complete fine-tuning steps take 77.67 and 80.70\,ms. Although the MoE models require fewer GFLOPs than the parameter-matched dense model, their measured latency is higher. Sparse expert execution introduces token dispatch and combination, irregular memory access, and multiple small expert-specific matrix multiplications, which increase kernel-launch and data-movement overhead and reduce GPU utilization compared with the large regular operations of a dense FFN.

\subsection{Representation Analysis Overhead}
\label{appx:analysis_overhead}
We measure the computational overhead of the representation analysis used to characterize layer-wise sharedness. The experiment trains a 12-layer dense Transformer for 512 steps with batch size eight, using the three self-supervised objectives adopted for pretraining.

Representations are collected every 16 training steps, resulting in 32 evaluation points. At each point, four fixed probe batches are used for each objective. The extracted representations are projected to 1024 dimensions before CKA and RSA.
\begin{table}[ht]
\centering
\caption{Wall-clock cost of the representation analysis. The total includes initialization, data loading, and visualization.}
\label{tab:analysis_overhead}
\begin{tabular}{lr}
\toprule
Stage & Time (s) \\
\midrule
Pilot training (512 steps) & 83.473 \\
Representation extraction & 21.954 \\
CKA and RSA computation & 77.394 \\
Visualization & 22.348 \\
Total & 214.341 \\
\bottomrule
\end{tabular}
\end{table}

The complete analysis takes 214.341\,s with a peak memory usage of 2175.5\,MiB. Pilot training and CKA/RSA computation take 83.473\,s and 77.394\,s, respectively, and constitute the main computational cost.

\section{Limitations and Future Work}\label{appx:limit}

First, while PRiSE-EEG scales to 125.58M parameters and 28,000 hours of data, this remains modest compared to the scale of contemporary language and vision foundation models. Furthermore, the inherent low signal-to-noise ratio of EEG data highlights a continual need for more advanced preprocessing techniques and self-supervised paradigms to learn truly robust representations.
Second, while our unified fine-tuning strategy is effective, integrating new tasks still requires retraining the model. A significant open challenge for the field is developing a truly universal EEG model that can generalize to novel tasks with zero- or few-shot prompting, similar to the capabilities of large language models.
Finally, our current work focuses on robust EEG representation; further research is needed to extend the model to multimodal scenarios such as Video-EEG, Image-EEG, Text-EEG, and fMRI-EEG for broader applicability.

\end{document}